\documentclass[twocolumn]{IEEEtran}
\usepackage[T1]{fontenc}
\usepackage[latin9]{inputenc}
\usepackage{color}
\usepackage{float}
\usepackage{textcomp}
\usepackage{amsmath}
\usepackage{amssymb}
\usepackage{graphicx}
\usepackage[unicode=true,
 bookmarks=true,bookmarksnumbered=true,bookmarksopen=true,bookmarksopenlevel=1,
 breaklinks=false,pdfborder={0 0 0},pdfborderstyle={},backref=false,colorlinks=false]
 {hyperref}
\hypersetup{pdftitle={Your Title},
 pdfauthor={Your Name},
 pdfpagelayout=OneColumn, pdfnewwindow=true, pdfstartview=XYZ, plainpages=false}

\makeatletter

\floatstyle{ruled}
\newfloat{algorithm}{tbp}{loa}
\providecommand{\algorithmname}{Algorithm}
\floatname{algorithm}{\protect\algorithmname}

\usepackage{cite}

\usepackage{algorithmic}
\usepackage{subfig}

\@ifundefined{showcaptionsetup}{}{%
 \PassOptionsToPackage{caption=false}{subfig}}
\usepackage{subfig}
\makeatother

\begin{document}
\title{Multi-User Pilot Pattern Optimization for Channel Extrapolation in
5G NR Systems}
\author{{\normalsize{}Yubo Wan, }\textit{\normalsize{}Graduate Student Member}{\normalsize{},}\textit{\normalsize{}
IEEE}{\normalsize{}, An Liu, }\textit{\normalsize{}Senior Member}{\normalsize{},}\textit{\normalsize{}
IEEE}{\normalsize{},}\textit{\normalsize{} }{\normalsize{}and Tony
Q. S. Quek, }\textit{\normalsize{}Fellow}{\normalsize{},}\textit{\normalsize{}
IEEE}{\normalsize{}}\thanks{This work was supported in part by National Key R\&D Program of China
(Grant No. 2021YFA1003304), in part by National Natural Science Foundation
of China under Grant 62071416, and in part by Zhejiang Provincial
Key Laboratory of Information Processing, Communication and Networking
(IPCAN), Hangzhou 310027, China. (Corresponding author: An Liu.)

Yubo Wan, An Liu are with the College of Information Science and Electronic
Engineering, Zhejiang University, Hangzhou 310027, China (email: wanyb@zju.edu.cn,
anliu@zju.edu.cn).\protect \\
\protect\hphantom{\protect\hphantom{}\protect\hphantom{}\protect\hphantom{}\protect\hphantom{}\protect\hphantom{}}~~~~T.
Q. S. Quek is with the Singapore University of Technology and Design,
Singapore 487372 (e-mail: tonyquek@sutd.edu.sg).}}
\maketitle
\begin{abstract}
Pilot pattern optimization in orthogonal frequency division multiplexing
(OFDM) systems has been widely investigated due to its positive impact
on channel estimation. In this paper, we consider the problem of multi-user
pilot pattern optimization for OFDM systems. In particular, the goal
is to enhance channel extrapolation performance for 5G NR systems
by optimizing multi-user pilot patterns in frequency-domain. We formulate
a novel pilot pattern optimization problem with the objective of minimizing
the maximum integrated side-lobe level (ISL) among all users, subject
to a statistical resolution limit (SRL) constraint. Unlike existing
literature that only utilizes ISL for controlling side-lobe levels
of the ambiguity function, we also leverage ISL to mitigate multi-user
interference in code-domain multiplexing. Additionally, the introduced
SRL constraint ensures sufficient delay resolution of the system to
resolve multipath, thereby improving channel extrapolation performance.
Then, we employ the estimation of distribution algorithm (EDA) to
solve the formulated problem in an offline manner. Finally, we extend
the formulated multi-user pilot pattern optimization problem to a
multiband scenario, in which multiband gains can be exploited to improve
system delay resolution. Simulation results demonstrate that the optimized
pilot pattern yields significant performance gains in channel extrapolation
over the conventional pilot patterns.
\end{abstract}


\section{Introduction}

Orthogonal frequency division multiplexing (OFDM) has been widely
recognized as an efficient modulation method in wireless communication
systems due to its high data rate transmission and its robustness
against frequency selectivity \cite{cimini1985analysis}. In time
division duplex (TDD) OFDM systems, the base station (BS) conducts
channel estimation based on the received uplink pilot symbols transmitted
from the user. The quality of the estimation hinges significantly
on the strategic placement of the pilots across subcarriers.

\begin{figure}[t]
\begin{centering}
\textsf{\includegraphics[width=8cm]{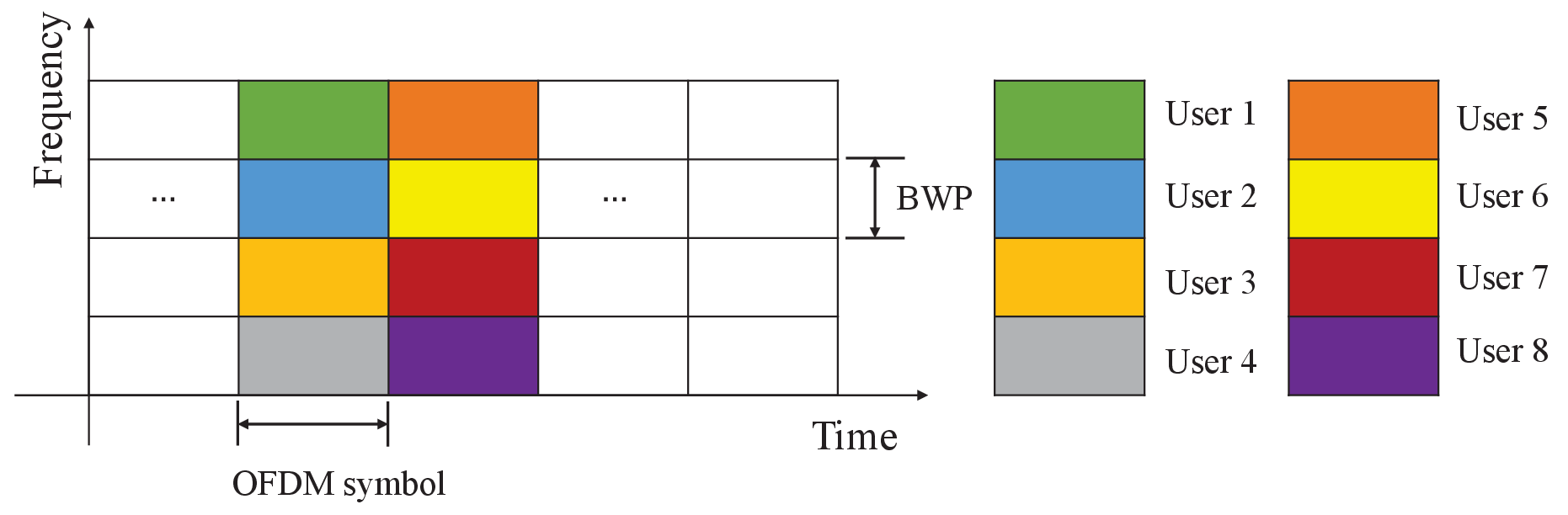}}
\par\end{centering}
\caption{\label{fig:Illustration-of-multiuser-1}An illustration of multi-user
frequency-domain and time-domain multiplexing.}
\end{figure}

In fifth-generation (5G) new radio (NR) systems, each user typically
allocates and transmits pilot symbols (also known as Sounding Reference
Signals (SRSs) in 5G NR systems) within a designated bandwidth part
(BWP) due to limitations in transmission power, occupying merely a
fraction of the overall system bandwidth for a given SRS period \cite{3gpp_5GNR}.
Additionally, 5G NR systems adopt an orthogonal multi-user SRS transmission
scheme employing three multiplexing methods: (1) \textbf{Code-Domain
Multiplexing}. Users transmit SRSs on the same time-frequency resource
using different Zadoff-Chu (ZC) pilot sequences \cite{3gpp_5GNR};
(2) \textbf{Frequency-Domain Multiplexing}. Users transmit SRSs on
distinct subcarriers within the same OFDM symbol; (3) \textbf{Time-Domain
Multiplexing}. Users transmit SRSs on different OFDM symbols within
the same BWP. These multiplexing methods are flexibly combined to
increase the number of users supported by the system, as shown \textcolor{black}{i}n
Fig. \ref{fig:Illustration-of-multiuser-1}.

Therefore, in 5G NR, the multi-user pilots transmission scheme merely
estimates a subset of the uplink channels in frequency-domain for
each user by utilizing received SRSs. This necessitates the application
of extrapolation method to deduce the remaining channels. However,
a new challenge arises: As compared to channel estimation, channel
extrapolation relies more heavily on high-resolution channel parameter
estimation to resolve the multipath components (MPCs) \cite{Extrapolation_1}.
With the current pilot pattern adopted in 5G NR, the capability for
channel extrapolation is constrained due to insufficient delay resolution
resulting from the limited bandwidth of a BWP, which represents the
bottleneck of current 5G NR systems. Numerous signal processing algorithms
for channel extrapolation have been proposed \cite{Extrapolation_2_MUSIC,Extrapolation_3_ESPRIT,Extrapolation_4_SBL,Extrapolation_9_multi,ChannelExtra_NR},
while research on pilot optimization to enhance channel extrapolation
performance remains underexplored, particularly in multi-user scenarios.

A substantial amount of research has been dedicated to pilot pattern
optimization in OFDM systems. The authors of \cite{Pilot_Simko} proposed
a pattern design by maximizing the channel capacity, which took channel
estimation errors into account. In \cite{Pilot_Zhang}, the pilot
pattern is designed based on the minimization of the bit error ratio
(BER). Moreover, a significant volume of research employed the mean
squared error (MSE) of a channel estimator as a key metric for pilot
pattern design, e.g., least squares (LS) estimator in \cite{Pilot_Barhumi,Pilot_Youssefi},
least absolute shrinkage and selection operator (LASSO) estimator
in \cite{Pilot_Chen}, and linear minimum mean squared error (LMMSE)
estimator in \cite{Pilot_Kim,Pilot_Sheng}. Notably, reference \cite{Pilot_Barhumi}
demonstrated that equi-powered and equi-spaced pilot-symbols lead
to the lowest MSE of LS estimator. Furthermore, in scenarios where
achieving an optimal pilot design is challenging due to time-varying
channels, the authors in \cite{Pilot_Youssefi} proposed an adaptive
pilot optimization scheme, yielding a suboptimal pilot arrangement.

Moreover, pilot optimization for sparse channel estimation has also
been extensively investigated driven by the emergence of compressed
sensing (CS) technology \cite{Pilot_Wang,Pilot_CS1,Pilot_CS2,Pilot_CS3}.
Minimizing mutual coherence of the measurement matrix has been widely
employed for pilot optimization, e.g., \cite{Pilot_Wang,Pilot_CS1,Pilot_CS3},
facilitating the exploitation of channel sparsity by CS-based algorithms.
Besides, the authors in \cite{Pilot_CS2} adopted the MSE of channel
estimation for sparse recovery algorithms as the optimization objective,
which is averaged over all possible channel data available at the
transmitter.

Nevertheless, several limitations are evident in the existing research:
\begin{itemize}
\item The majority of studies concentrate on single-user scenarios, where
the designed pilot pattern may not be directly applicable to practical
multi-user scenarios. Furthermore, in real communication systems such
as 5G NR, code-domain multiplexing has to be considered for supporting
more users. In multi-user scenarios, e.g., \cite{Pilot_Hayder}, however,
the impact of pilot pattern on code-domain multiplexing of the systems
has not been taken into account.
\item Few studies explore pilot pattern optimization with the aim of enhancing
multipath resolution capability of the systems for channel extrapolation.
As explained before, channel extrapolation plays a crucial role in
5G NR systems and insufficient multipath resolution hinders channel
extrapolation performance. However, most references pay attention
to improve channel estimation accuracy through pilot optimization,
but ignore the capability of the system to resolve MPCs. Particularly,
the MSE of LS estimator and LMMSE estimator has been widely used for
pilot pattern design, e.g., \cite{Pilot_Barhumi,Pilot_Youssefi,Pilot_Kim,Pilot_Sheng}.
Though LS and LMMSE estimators estimate the channel on the BWP that
transmits SRS with high accuracy under the specially optimized pilot
patterns, the extrapolation range of them is very limited \cite{Extrapolation_1}.
\item They mainly rely on the prior information of channel, e.g., realization
of channel samples \cite{Pilot_CS2,Pilot_Chen}, or precise channel
statistical information \cite{Pilot_Zhang,Pilot_Kim,Pilot_Sheng}.
However, acquiring such prior information is difficult in practical
systems.
\item Multiband technology has been widely employed in wireless systems,
which combines multiple non-contiguous frequency bands to achieve
high-resolution multipath delay estimation \cite{ESPRIT2,TSGE_early}.
Nonetheless, existing pilot pattern optimization in frequency-domain
is typically based on a single contiguous frequency band, which cannot
be applied to multiband scenarios.
\end{itemize}

Motivated by the limitations of existing pilot optimization methods
and multi-user pilots transmission scheme employed in 5G NR, this
paper proposes a novel scheme for the design of multi-user pilot pattern
in uplink OFDM systems. We formulate a multi-user pilot pattern optimization
problem and employ an evolutionary algorithm called estimation of
distribution algorithm (EDA) \cite{Pilot_Wang} to solve the formulated
problem. This scheme addresses the issue of insufficient extrapolation
capability in 5G NR systems and can be implemented in an offline manner.
The optimized pilot pattern holds the potential to reduce multi-user
interference caused by code-domain multiplexing and enhance the delay
resolution of the system, thereby improving the overall channel extrapolation
performance.

The main contributions of this paper are summarized as follows.
\begin{enumerate}
\item We formulate a novel pilot pattern optimization problem tailored for
multi-user OFDM systems. The multi-user pilots are designed to minimize
the worst integrated side-lobe level (ISL) \cite{ISL} among users
under the multi-user statistical resolution limit (SRL) constraint
\cite{SRL}. The performance metric, ISL, is derived from delay ambiguity
function (AF) and can control the relative amount of energy in side-lobe
portions with respect to the main-lobe portions of AF. As compared
to existing methods that optimize the performance metrics such as
MSE of channel estimation \cite{Pilot_Chen,Pilot_Kim,Pilot_Sheng},
minimizing ISL offers several advantages: (1) Enhanced delay estimation
accuracy due to effective side-lobe suppression to mitigate the delay
ambiguity \cite{ISL2}; (2) Improved multi-user interference cancellation
capability in multi-user code-domain multiplexing, as explained in
Subsection \ref{subsec:ISL}; (3) ISL optimization does not rely on
any channel prior information. Then, the introduced SRL is a performance
bound that provides a delay resolution limit for any practical method.
The SRL constraint guarantees sufficient delay resolution of the optimized
systems for channel extrapolation. The proposed optimization problem
has the advantage of allowing offline optimization without prior information
of the channel required.
\item The formulated multi-user pilot pattern optimization problem is extended
to multiband scenarios, where the pilots are optimized across non-contiguous
frequency bands/subcarriers. As compared to single-band scenarios,
pilot optimization in multiband scenarios has the potential to enhance
the delay resolution, aligning with our goal to improve channel extrapolation
performance. Based on a formulated multiband signal model, we derive
the corresponding ISL and SRL metrics in multiband multi-user scenarios
and finally formulate the multiband multi-user pilot pattern optimization
problem.
\item Extensive optimization results are presented in both single-band and
multiband scenarios, considering the impact of multi-user interference
on the system with both frequency-domain multiplexing and code-domain
multiplexing. Particularly, the optimized pilot patterns have been
presented, which can provide useful guidance for the pilot design
in 5G NR systems. Superior channel extrapolation performance based
on the optimized pilot pattern has been validated using realistic
channels, i.e., the urban macro (UMa) scenario depicted in 3GPP R16
specifications \cite{3gpp_Rel16}.
\end{enumerate}

The rest of this paper is organized as follows. In Section \ref{sec:System-Model},
we describe the system and signal model. In Section \ref{sec:EST},
we firstly introduce the performance metrics ISL and SRL. Then, multi-user
pilot pattern optimization problem is formulated and the adopted EDA
algorithm to solve the formulated problem is introduced. In Section
\ref{sec:Tra}, the formulated problem is extended from single-band
to multiband scenarios. Finally, simulation results and conclusions
are given in Sections \ref{sec:Simulation-Results} and \ref{sec:Conclusion},
respectively.

\textit{Notations:} The notation $\left\Vert \cdot\right\Vert _{2}$
denotes the $\ell_{2}$-norm, $\mathrm{diag}\left(\cdot\right)$ constructs
a diagonal matrix from its vector argument, and $\ast$ denotes the
Hadamard product. The transpose, conjugate transpose, and inverse
are denoted by $(\cdot)^{T},(\cdot)^{H},(\cdot)^{-1}$ respectively.
The notation $\mathcal{N}(\mathrm{\mu},\Sigma)$ denotes a Gaussian
normal distribution with mean $\mu$ and variance $\Sigma$, and $\mathbb{E}_{\mathbf{z}}[\cdot]$
denotes the expectation operator with respect to the random vector
$\mathbf{z}$.

\section{System and Signal Model\label{sec:System-Model}}

\subsection{System Model}

Consider a multi-user OFDM system that adopts both code-domain multiplexing
and frequency-domain multiplexing to support multi-user uplink transmission,
as illustrated in Fig. \ref{sec:System-Model}. In the considered
system, each single antenna user transmits uplink SRSs to the BS.
Note that the extension of our scenario to a user with multiple antennas
is trivial, since the antennas in the same user are generally assigned
with orthogonal SRSs. We employ the ZC sequences as the transmitted
pilot symbols, which have been adopted in 5G NR and specified in 3GPP
5G NR Release 16 \cite{3gpp_5GNR}. ZC sequences exhibit the useful
property that cyclically shifted versions of themselves are orthogonal
to one another. Specifically, the SRS sequence is generated by a cyclic
shift of a base sequence, which is given by
\begin{equation}
\mathrm{r}_{u}^{\gamma}\left(n\right)=e^{j\gamma n}\bar{\mathrm{r}}_{u}\left(n\right),\label{eq:SRS generation}
\end{equation}

\noindent where $\bar{\mathrm{r}}_{u}\left(n\right),n=0,...,N-1,$
is the base sequence, $N$ denotes the number of subcarriers in full
frequency band, $u$ denotes the index number of the base sequence.
The base sequence is cyclically shifted by the term $e^{j\gamma n}$
to generate orthogonal sequences, which means that the inner product
of $\mathrm{r}_{u}^{\gamma}\left(n\right)$ and $\bar{\mathrm{r}}_{u}\left(n\right)$
equals to zero. Generally, the pilot sequences are chosen from the
orthogonality set $\Gamma=\left\{ \bar{\mathrm{r}}_{u}\left(n\right),\mathrm{r}_{u}^{\gamma}\left(n\right)\right\} ,\alpha\in\mathbb{Z}$,
to reduce the multi-user interference. However, the multi-user interference
in code-domain multiplexing transmission not only relies on the orthogonality
of the SRS sequences, but also relies on the multi-user pilot patterns
in frequency-domain, which motivates the utilization of ISL to reduce
multi-user interference in pilot pattern optimization, as detailed
in Subsection \ref{subsec:ISL}.

\begin{figure}[htbp]
\begin{centering}
\includegraphics[width=8cm]{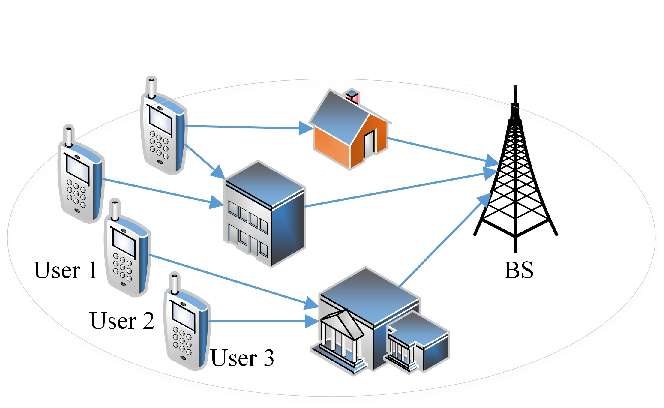}
\par\end{centering}
\caption{\label{fig:system_model}Illustration of multi-user uplink system.}
\end{figure}

\subsection{Pilot Signal Model}

We consider $G$ user groups with $G$ pilot pattern vectors, $\mathbf{w}_{1},...,\mathbf{w}_{G}$,
and the users in the $g$-th group share the same pilot pattern vector
$\mathbf{w}_{g}$ but adopt different SRS sequences for supporting
code-domain multiplexing. For the $(g,z)$-th user, i.e., the $z$-th
user of the $g$-th user group, the channel frequency response (CFR)
is given by
\begin{flalign}
\mathbf{h}_{g,z} & =\sum_{k=1}^{K_{g,z}}\alpha_{g,z,k}\boldsymbol{a}\left(\tau_{g,z,k}\right),\label{eq:H channel model F}
\end{flalign}
where
\[
\boldsymbol{a}(\tau_{g,z,k})=\left[1,e^{-j2\pi f_{s}\tau_{g,z,k}},...,e^{-j2\pi(N-1)f_{s}\tau_{g,z,k}}\right]^{T}
\]
denotes frequency-domain steering vector, $\mathbf{h}_{g,z}\in\mathbb{C}^{N\times1}$,
$K_{g,z}$ is the number of MPCs, $f_{s}$ denotes the subcarrier
spacing, $\alpha_{g,z,k}$ and $\tau_{g,z,k}$ denote the complex
path gain and time delay of the $k$-th path, respectively. As our
focus is on pilot pattern optimization in frequency-domain, which
primarily impacts delay estimation performance, and considering that
the different antennas of the BS receive equivalent training pilot
sequences transmitted from the users, we consider CFR from a single
antenna of the BS for simplicity.

In the pilot training phase, all users synchronously send their pilot
signals to the BS. Then, the received signals in frequency-domain
can be represented as
\begin{alignat}{1}
\mathbf{y} & =\sum_{g=1}^{G}\sum_{z=1}^{Z}\textrm{diag}\left(\mathbf{w}_{g}\ast\mathbf{x}_{z}\right)\mathbf{h}_{g,z}+\mathbf{n}_{g,z},\label{eq:recF}
\end{alignat}
where $\mathbf{y}\in\mathbb{C}^{N\text{\texttimes}1}$, $\mathbf{x}_{z}\in\mathbb{C}^{N\text{\texttimes}1},z=1,...,Z,$
denotes the $z$-th SRS sequence generated from (\ref{eq:SRS generation}),
and $\mathbf{n}_{g,z}\in\mathbb{C}^{N\text{\texttimes}1}$ denotes
the additive white Gaussian noise (AWGN) with each element having
zero mean and variance $\sigma_{e,g,z}^{2}$. We denote the number
of non-zero elements of the pilot pattern vector $\mathbf{w}_{g}\in\left\{ 0,1\right\} ^{N\times1}$
as $P_{g}$. The non-zero elements in $\mathbf{w}_{g}$ indicate the
location of the transmitted SRS symbols of the $g$-th group in subcarriers
and $P_{g}$ represents the length of transmitted SRS sequence of
the $g$-th group.

\section{Multi-User Pilot Pattern Optimization\label{sec:EST}}

In this section, we firstly elaborate the performance metrics ISL
and SRL used in the multi-user pilot pattern optimization scheme.
Then, we formulate the optimization problem and employ the EDA algorithm
to solve this problem.

\vspace{-0.15in}

\subsection{Performance Metrics\label{subsec:Performance-Metrics}}

\subsubsection{ISL\label{subsec:ISL}}

The normalized ISL is defined as \cite{ISL}
\begin{equation}
\textrm{ISL}\triangleq\frac{\int_{\mathcal{R}_{s}}\left|\chi(\Delta\tau)\right|^{2}\mathrm{d}\tau}{\left|\mathcal{R}_{s}\right|\left|\chi(0)\right|^{2}},\label{eq:ISL}
\end{equation}
where $\chi(\Delta\tau)$ denotes the AF of the OFDM waveform at a
given delay mismatch $\Delta\tau\triangleq\tau_{2}-\tau_{1}$, $\mathcal{R}_{s}$
denotes the side-lobe region in the delay domain of the AF. Recalling
the observation model in (\ref{eq:recF}), the AF of the $(g,z)$-th
user can be derived as \cite{ISL}
\begin{eqnarray}
\chi_{g,z}(\Delta\tau) & = & \left(\mathbf{w}_{g}\ast\mathbf{x}_{z}\ast\boldsymbol{a}(\tau_{1})\right)^{H}\left(\mathbf{w}_{g}\ast\mathbf{x}_{z}\ast\boldsymbol{a}(\tau_{2})\right)\nonumber \\
 & = & \left[\left(\mathbf{w}_{g}\ast\mathbf{w}_{g}\right)\ast\left(\mathbf{x}_{z}^{*}\ast\mathbf{x}_{z}\right)\right]^{T}\left[\boldsymbol{a}^{*}(\tau_{1})\ast\boldsymbol{a}(\tau_{2})\right]\nonumber \\
 & = & \mathbf{w}_{g}^{T}\boldsymbol{a}(\Delta\tau).\label{eq:AF}
\end{eqnarray}
As can be seen, the AF is independent of the ZC sequences and thus
we omit the subscript $z$ as $\chi_{g}(\Delta\tau)$ for the $g$-th
group. From the definition of ISL, we can know that it characterizes
the side-lobe levels of AF and does not rely on any prior information
of the channel. This performance metric has been widely employed for
waveform design in radar systems \cite{ISL,ISL2,ISL_2} to lower the
side-lobe level. However, to the best of our knowledge, utilizing
ISL to reduce multi-user interference in code-domain multiplexing
for multi-user pilot transmission systems has not been studied yet.

Specifically, the multi-user system with code-domain multiplexing
mitigates the multi-user interference by decoupling the received signal
$\mathbf{y}$ for each user in delay domain, which contains three
steps:

(1) Multiply $\mathbf{y}$ with the conjugation of the SRS sequence
under the corresponding pilot pattern, e.g., for the $(g,z)$-th user,
$\tilde{\mathbf{y}}_{g,z}=\textrm{diag}\left(\mathbf{w}_{g}\ast\mathbf{x}_{z}^{*}\right)\mathbf{y}$;

(2) Transform $\tilde{\mathbf{y}}_{g,z}$ into the delay domain by
multiplying an inverse discrete Fourier transform (IDFT) matrix $\mathbf{F}$,
i.e., $\tilde{\mathbf{y}}_{g,z}^{D}=\mathbf{F}\tilde{\mathbf{y}}_{g,z}$;

(3) Remove delays that are out of the delay spread region, i.e., $\hat{\mathbf{y}}_{g,z}^{D}=\mathcal{F}\left\{ \tilde{\mathbf{y}}_{g,z}^{D}\right\} $.

\begin{figure}[t]
\centering{}\textcolor{blue}{\includegraphics[width=8cm]{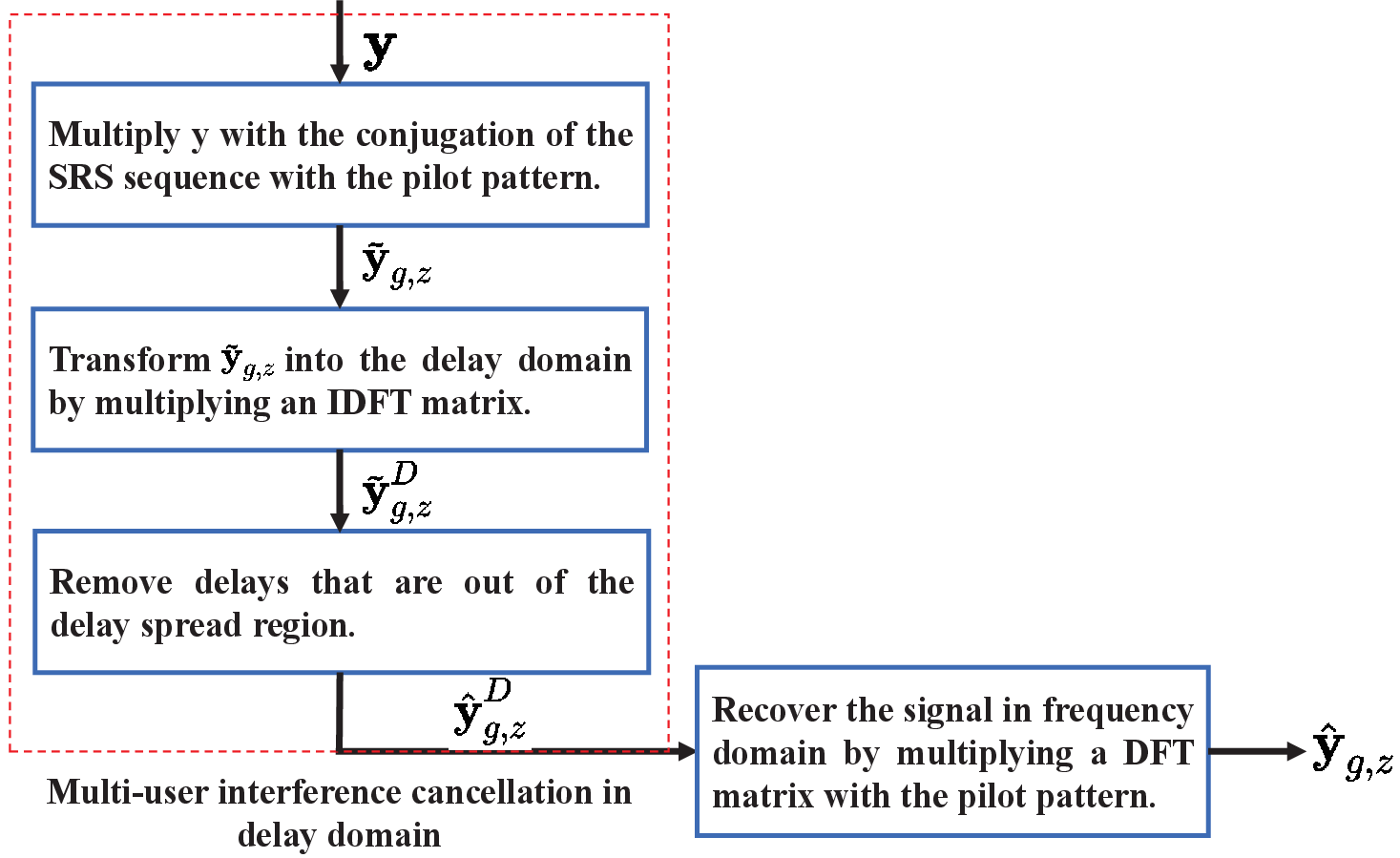}\caption{\label{fig:FlowChart}The flow chart of the multi-user interference
cancellation procedure.}
}
\end{figure}

A flow chart depicting the multi-user interference cancellation procedure
is shown in Fig. \ref{fig:FlowChart}. Note that though the original
SRS sequences of two users may be orthogonal, the multi-user interference
in code-domain multiplexing cannot be totally eliminated due to the
impact of pilot patterns. Hence, after the above process of interference
cancellation in delay domain, the multi-user interference is only
partially eliminated, which motivates the utilization of ISL to design
pilot pattern to further mitigate multi-user interference. The rationale
behind the benefit of multi-user interference cancellation for code-domain
multiplexing from the optimization of ISL is reasonable, since the
cancellation of multi-user interference exactly occurs in delay domain,
which aligns with the domain where ISL is defined. \textcolor{blue}{}

\begin{figure}[t]
\begin{centering}
\begin{minipage}[t]{0.45\textwidth}%
\begin{center}
\subfloat[]{\begin{centering}
\includegraphics[width=80mm]{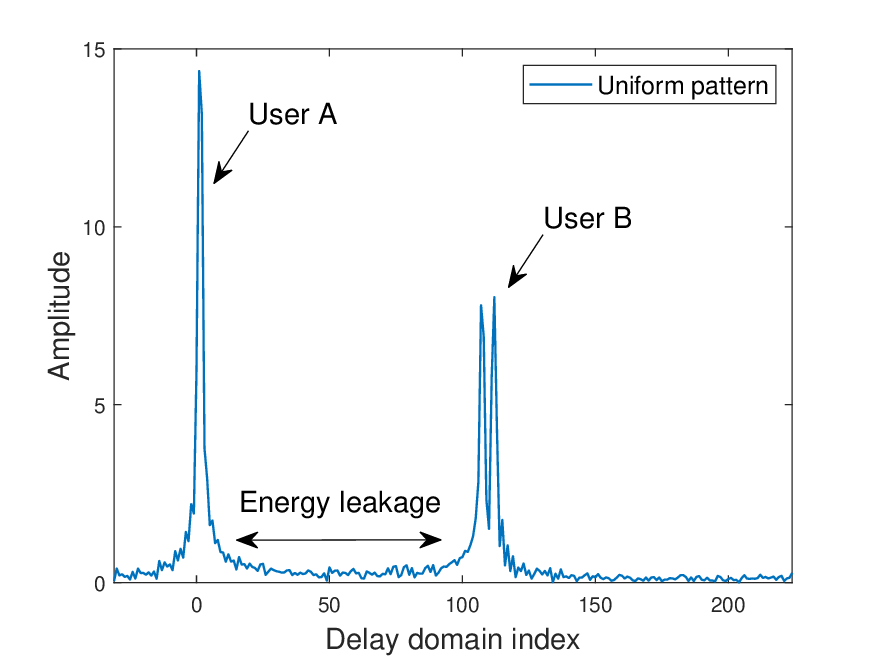}
\par\end{centering}
}
\par\end{center}%
\end{minipage}\hfill{}%
\begin{minipage}[t]{0.45\textwidth}%
\begin{center}
\subfloat[]{\begin{centering}
\includegraphics[width=8cm]{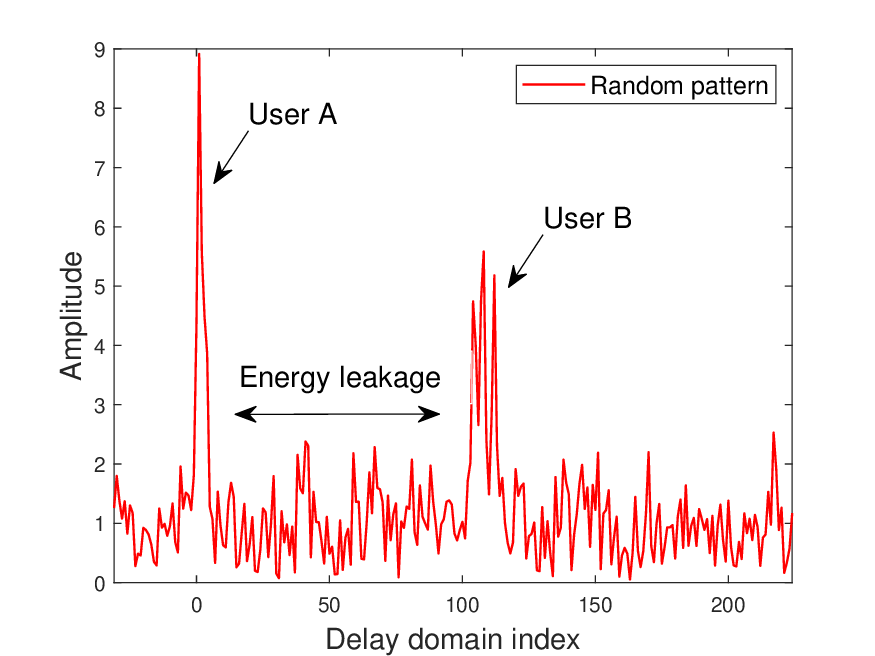}
\par\end{centering}
}
\par\end{center}%
\end{minipage}
\par\end{centering}
\medskip{}

\centering{}\caption{\label{fig:ISL_Sec_Delay_Profile}An illustration of magnitude curves
of a user group in delay domain.}
\end{figure}

\begin{figure}[t]
\centering{}\textcolor{blue}{\includegraphics[width=8cm]{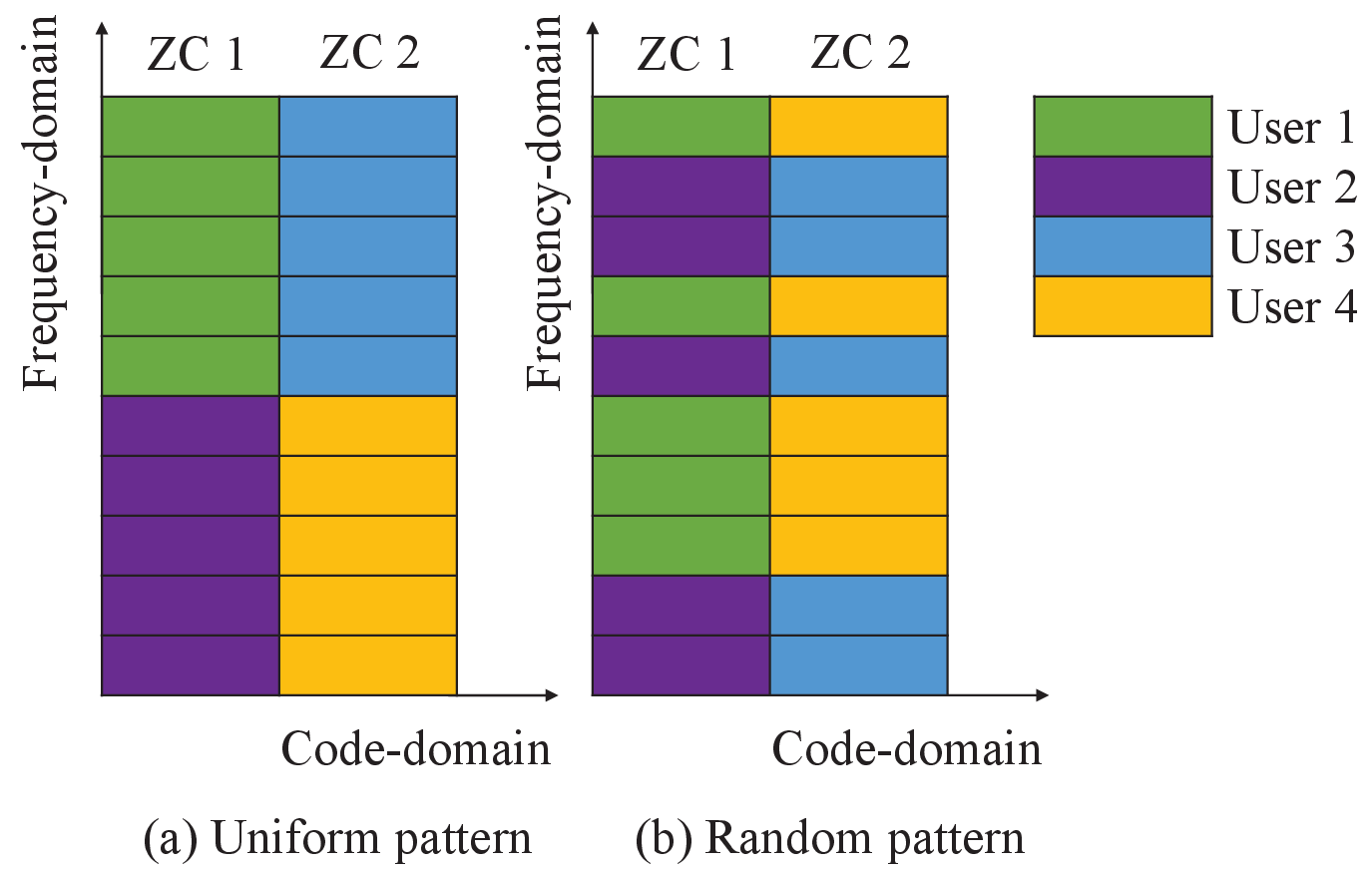}\caption{\label{fig:Bas_pattern}An illustration of uniform pilot pattern and
random pilot pattern.}
}
\end{figure}

To demonstrate the indication of ISL on interference cancellation,
we present the magnitude curves of $\tilde{\mathbf{y}}_{g,z}^{D}$
in Fig. \ref{fig:ISL_Sec_Delay_Profile} under two conventional pilot
patterns, i.e., uniform pattern and random pattern, as illustrated
in Fig. \ref{fig:Bas_pattern}, where we set 4 users being divided
into $G=2$ user groups and different user groups occupy different
subcarriers. The interference between the two users in the same group
is handled by code-domain multiplexing. We set $P_{g}=128,\forall g,$$N=256,$$f_{s}=120$
KHz. The channel has 2 paths with delays uniformly generated from
$[0,400]$ ns for all users. The signal-to-noise ratios (SNR) is set
as 15 dB. The averaged ISL value among users based on the uniform
pattern and random pattern is $-23$ dB and $-21$ dB, respectively.
As can be seen, the pattern scheme with lower ISL has less energy
leakage from the main-lobe and thus has lower side-lobe level. Moreover,
there is larger multi-user interference for code-domain multiplexing
due to energy leakage from user B to the delay spread region of user
A for the random pilot pattern scheme due to the higher ISL.

Finally, we can recover the signal for each user in frequency-domain
by multiplying a discrete Fourier transform (DFT) matrix with the
corresponding pilot pattern as $\hat{\mathbf{y}}_{g,z}=\mathbf{w}_{g}\ast(\mathbf{F}^{H}\hat{\mathbf{y}}_{g,z}^{D})$.
The normalized mean square error (NMSE) (as defined in (\ref{eq:NMSE}))
of the recovered channel in frequency domain for uniform pattern and
random pattern scheme is $0.0085$ and $0.3361$, respectively. As
indicated by the ISL, the uniform pattern scheme is less affected
by the multi-user interference than the random pattern scheme, and
thus leads to a better channel estimation performance.
\textcolor{blue}{}

Then, since ISL is also independent of the SRS sequence $\mathbf{x}_{z}$
as the AF, we derive the closed-form expression of the ISL of the
$g$-th group. Substituting (\ref{eq:AF}) into (\ref{eq:ISL}), the
ISL of the $g$-th group can be written as
\begin{equation}
\textrm{ISL}_{g}=\frac{\mathbf{w}_{g}^{T}\mathbf{G}\mathbf{w}_{g}}{\left|\mathcal{R}_{s}\right|\left(\mathbf{w}_{g}^{T}\boldsymbol{1}\right)^{2}},\label{eq:ISL_2}
\end{equation}
where
\begin{equation}
\mathbf{G}=\int_{\mathcal{R}_{s}}\boldsymbol{a}(\Delta\tau)\boldsymbol{a}^{H}(\Delta\tau)\mathrm{d}\tau.
\end{equation}
For a symmetric interval $\mathcal{R}_{s}\triangleq\left[-b,-a\right]\cup\left[a,b\right],a,b\in\mathbb{R^{\textrm{+}}}$,
$\mathbf{G}$ is a symmetric Toeplitz matrix and hence can be described
by its first column $\boldsymbol{f}(n),n=0,...,N-1$, where
\[
\boldsymbol{f}(n)=\left\{ \begin{array}{cc}
\frac{\sin(2\pi nf_{s}b)-\sin(2\pi nf_{s}a)}{\pi nf_{s}}\, & n=1,...,N-1,\\
2b-2a & n=0.
\end{array}\right.
\]

\subsubsection{Delay SRL}

The performance metric delay SRL is defined as \cite{SRL}
\begin{eqnarray}
 & \mathrm{SRL} & \triangleq\Delta\tau\nonumber \\
\text{\textrm{s.t.}} & \Delta\tau & =\sqrt{\textrm{CRB}_{\Delta\tau}},\label{eq:SRL}
\end{eqnarray}
where $\textrm{CRB}_{\Delta\tau}$ denotes the Cram�r-Rao bound (CRB)
for $\Delta\tau$. CRB serves as a MSE lower bound for the unbiased
estimator \cite{sengijpta1995fundamentals}. The delay SRL is defined
using the CRB of signal parameters for the cases where two path delays
are closely located. According to the definition in (\ref{eq:SRL}),
the delay SRL is the delay separation that is equal to its own root
squared CRB. If there are multiple solutions in (\ref{eq:SRL}), the
SRL can be defined as the smallest delay separation among all the
solutions. In this definition, the delays can be exactly \textquotedblleft resolved\textquotedblright{}
when the standard deviation of the delay separation estimation is
equal to the true separation. Delay SRL provides the highest delay
resolution achievable by any unbiased estimator.

Then, the CRB for the delay separation of the $(g,z)$-th user, $\textrm{CRB}_{\Delta\tau_{g,z}}$,
is derived as
\begin{eqnarray}
\textrm{CRB}_{\Delta\tau_{g,z}} & = & \frac{\partial\Delta\tau_{g,z}}{\partial\boldsymbol{\theta}_{g,z}}^{T}\mathbf{J}_{\boldsymbol{\theta}_{g,z}}^{-1}\frac{\partial\Delta\tau_{g,z}}{\partial\boldsymbol{\theta}_{g,z}},\label{eq:C_delattau-1}\\
 & = & \mathbf{J}_{\boldsymbol{\theta}_{g,z}}^{-1}(1,1)+\mathbf{J}_{\boldsymbol{\theta}_{g,z}}^{-1}(2,2)-\mathbf{J}_{\boldsymbol{\theta}_{g,z}}^{-1}(1,2)\nonumber \\
 &  & -\mathbf{J}_{\boldsymbol{\theta}_{g,z}}^{-1}(2,1),\nonumber 
\end{eqnarray}
where $\mathbf{J}_{\boldsymbol{\theta}_{g,z}}\triangleq\mathbb{E}_{\mathbf{y}}\left[-\frac{\partial^{2}\ln\mathit{f}(\mathbf{y}|\boldsymbol{\theta}_{g,z})}{\partial\boldsymbol{\theta}_{g,z}\partial\boldsymbol{\theta}_{g,z}^{T}}\right]$
is the Fisher information matrix (FIM) associated with the vector
$\boldsymbol{\theta}_{g,z}$, $f(\mathbf{y}|\boldsymbol{\theta}_{g,z})$
is the likelihood function of the random vector $\mathbf{y}$ conditioned
on $\boldsymbol{\theta}_{g,z}$, and $\boldsymbol{\theta}_{g,z}\triangleq[\boldsymbol{\tau}_{g,z}^{T},\boldsymbol{\alpha}_{g,z}^{T}]$
consisting of the unknown parameters:
\begin{equation}
\begin{aligned} & \boldsymbol{\tau}_{g,z}=\left[\tau_{g,z,1},\tau_{g,z,2}\right]^{T},\\
 & \boldsymbol{\alpha}_{g,z}=\left[(\boldsymbol{\alpha}_{g,z}^{R})^{T},(\boldsymbol{\alpha}_{g,z}^{I})^{T}\right]^{T},\\
 & \boldsymbol{\alpha}_{g,z}^{R}=\left[\alpha_{g,z,1}^{R},\alpha_{g,z,2}^{R}\right]^{T},\\
 & \boldsymbol{\alpha}_{g,z}^{I}=\left[\alpha_{g,z,1}^{I},\alpha_{g,z,2}^{I}\right]^{T},
\end{aligned}
\end{equation}
in which $\boldsymbol{\alpha}_{g,z}^{R}$ and $\boldsymbol{\alpha}_{g,z}^{I}$
denote the real and imaginary parts of $\boldsymbol{\alpha}_{g,z}$,
respectively. We set $K_{g,z}=2$ since delay SRL is exactly defined
based on two close delays. Note that in practical scenarios with the
number of multipath $K_{g,z}>2$, the pilot pattern results of the
adopted offline optimization can still be effective to improve the
channel extrapolation performance, as verified by the numerical simulations
in Section \ref{sec:Simulation-Results}. The reader is referred to
Appendix \ref{subsec:AppexA} for detailed derivations of $\mathbf{J}_{\boldsymbol{\theta}_{g,z}}$.

\subsection{Problem Formulation}

Generally, the uniform pilot pattern leads to a small ISL value, while
the random pilot pattern leads to a small SRL value. However, channel
extrapolation for multi-user requires the ISL to be as small as possible
and the SRL cannot be too large. Therefore, we aim to minimize the
maximum ISL among all users by pilot pattern optimization to improve
the overall multi-user system performance. At the same time, a relatively
small SRL should be guaranteed. Consequently, the problem can be formulated
as
\begin{eqnarray}
\mathcal{P}: & \underset{\mathcal{X}}{\min} & \underset{g}{\max}\left\{ \textrm{ISL}_{g}\right\} \nonumber \\
 & \text{s.t. } & \ensuremath{\textrm{SRL}_{g,z}\leq\beta_{g,z},\forall g,z},\label{eq:P2_SRL}\\
 &  & \sum_{g=1}^{G}\mathbf{w}_{g}\leq\boldsymbol{1},\label{eq:P2_bandwidth}
\end{eqnarray}
where $\mathcal{X}\triangleq\left\{ \mathbf{w}_{1},...,\mathbf{w}_{G}\right\} $
denotes the optimization variables set, $\textrm{SRL}_{g,z}$ denotes
the SRL of the $(g,z)$-th user, and $\beta_{g,z}$ denotes the upper
limit on the SRL of the $(g,z)$-th user. Minimizing ISL leads to
less multi-user interference in code-domain and less delay ambiguity
for delay estimation since the side-lobe of the AF is suppressed.
Constraint (\ref{eq:P2_SRL}) guarantees sufficient delay resolution
of the multi-user system for channel extrapolation. Consequently,
the multi-user system is expected to reach an enhanced channel extrapolation
performance under the optimized multi-user pilot pattern of problem
$\mathcal{P}$. Constraint (\ref{eq:P2_bandwidth}) ensures that the
pilot symbols allocated to different users are not overlapped in subcarriers.

Note that the problem $\mathcal{P}$ is optimized in an offline manner,
which does not require any instantaneous realization of channel samples
or prior information of channel statistical characteristics as other
pilot pattern optimization schemes, e.g., \cite{Pilot_CS2,Pilot_Chen,Pilot_Zhang,Pilot_Kim,Pilot_Sheng}.
Specifically, since the calculation of delay SRL requires true value
of the noise variance and the path gains, we set the noise variance
and the path gains of the users in the same group to be the same as
a fixed value, i.e., $\sigma_{e,g,1}=\sigma_{e,g,2}=...=\sigma_{e,g,Z}=\sigma_{g},\forall g$,
$\alpha_{g,1,k}=\alpha_{g,2,k}=...=\alpha_{g,Z,k}=\rho_{g,k},\forall g,k,$
so that the users in the same group have the same SRL and the SRL
can be readily calculated in an offline manner. Note that the setting
value of $\sigma_{g}$ and $\rho_{g,k}$ have relatively little effect
on the final channel extrapolation performance since we can adjust
the corresponding $\beta_{g,z}$ to ensure that the delay SRL constraint
(\ref{eq:P2_SRL}) is always effective to shape the pilot pattern.
In other words, the optimal solution for $\mathcal{P}$ under a typical
value of $\sigma_{g},\rho_{g,k},\beta_{g,z}$ can also achieve a pretty
good channel extrapolation performance under a wide range of $\sigma_{g},\rho_{g,k},\beta_{g,z}$.
In practice, we set $\sigma_{g},\rho_{g,k}$ to a typical SNR of $15$
dB (In the simulations, we have set $\rho_{g,1}=\rho_{g,2}=1$, $\sigma_{g}=0.1778,\forall g$).
For given $\sigma_{g}$ and $\rho_{g,k}$, we calculate the SRL value
under a random pilot pattern as a reference for setting $\beta_{g,z}$.
It is because random pilot pattern generally leads to a much better
SRL than other pilot pattern schemes. In Section \ref{sec:Simulation-Results},
the effectiveness of our proposed offline pilot pattern optimization
scheme has been verified in a practical multipath channel model, which
outperforms the benchmark schemes with significant performance gains.

Finally, substituting (\ref{eq:ISL_2}) and (\ref{eq:SRL}) into $\mathcal{P}$,
the problem can be equivalently formulated as
\begin{eqnarray*}
\mathcal{P}: & \underset{\mathcal{X}}{\min} & \underset{g}{\max}\left\{ \frac{\mathbf{w}_{g}^{T}\mathbf{G}\mathbf{w}_{g}}{\left|\mathcal{R}_{s}\right|\left(\mathbf{w}_{g}^{T}\boldsymbol{1}\right)^{2}}\right\} \\
 & \text{s.t. } & \Delta\tau_{g}\leq\beta_{g},\forall g,\\
 &  & \ensuremath{\sqrt{\textrm{CRB}_{\Delta\tau_{g}}}=\Delta\tau_{g}},\forall g,\\
 &  & \sum_{g=1}^{G}\mathbf{w}_{g}\leq\boldsymbol{1}.
\end{eqnarray*}
Note that we omit the subscript $z$ in the delay SRL constraint since
the problem $\mathcal{P}$ is independent of the subscript $z$ in
offline optimization. Pilot pattern optimization of $\mathcal{P}$
is an integer nonlinear programming (INLP) problem. Finding the optimal
solution of an INLP problem is intractable because it is generally
non-deterministic polynomial-time hard (NP-hard). In next subsection,
we will employ the EDA method to solve this problem.

\subsection{EDA for Solving $\mathcal{P}$}

EDA is an evolutionary algorithm that operates by learning and sampling
the probability distribution of the best individuals within a population
at each generation. Unlike producing a specific individual, EDA introduces
population diversity by sampling individuals from the probability
distribution. This characteristic endows EDA with a stronger global
search ability and robustness compared to other evolutionary algorithms
when solving INLP problems \cite{Pilot_Wang}.

Generally, EDA starts with a random initialization of a population
$\mathcal{W}^{(0)}$ composed of $Q$ individuals, i.e., $\mathcal{W}^{(0)}=\{\mathbf{W}_{1}^{(0)},...,\mathbf{W}_{Q}^{(0)}\}$,
where $\mathbf{W}_{q}^{(0)}\in\left\{ 0,1\right\} ^{N\times G}$ is
a binary matrix representing a candidate of the multi-user pilot pattern
matrix, $\mathbf{W}\triangleq[\mathbf{w}_{1},...,\mathbf{w}_{G}]\in\left\{ 0,1\right\} ^{N\times G}$.
Each row of $\mathbf{W}_{q}^{(0)}$ has at most one non-zero element
in order to satisfy constraint (\ref{eq:P2_bandwidth}). In addition,
only the samples satisfying constraint (\ref{eq:P2_SRL}) are selected
as the candidate individuals. The SRL is calculated by one-dimensional
search of $\Delta\tau$ in (\ref{eq:SRL}) with acceptable computational
complexity.

Then, for the $i$-th iteration, we select the best $T$ $(T<Q)$
individuals from $\mathcal{W}^{(i-1)}$ denoted as $\mathcal{\widetilde{W}}^{(i)}=\{\widetilde{\mathbf{W}}_{1}^{(i)},...,\widetilde{\mathbf{W}}_{T}^{(i)}\}$.
The goodness of the individuals is measured by the objective function
value in $\mathcal{P}$ (also called fitness in EDA algorithms). In
our problem, a smaller fitness (ISL value) indicates a better individual.

Next, we calculate the probability of a pilot symbol occurring on
the $n$-th subcarrier of the $g$-th group by averaging among the
individuals in $\mathcal{\widetilde{W}}^{(i)}$, i.e.,
\begin{equation}
p^{(i)}(n,g)=\frac{1}{T}\sum_{t=1}^{T}\widetilde{\mathbf{W}}_{t}^{(i)}(n,g).\label{eq:Pro}
\end{equation}

Then, we generate a new population $\mathcal{W}^{(i)}$ with individuals
$\mathbf{W}_{q}^{(i)}(n,g)\sim\textrm{Bern}(p^{(i)}(n,g))$, where
$\textrm{Bern}(p^{(i)}(n,g))$ represents a Bernoulli distribution
with probability $P(\mathbf{W}_{q}^{(i)}(n,g)=1)=p^{(i)}(n,g)$, under
the condition of satisfying constraints (\ref{eq:P2_SRL})-(\ref{eq:P2_bandwidth}).
Besides, we preserve the best individual from $\mathcal{W}^{(i-1)}$
to $\mathcal{W}^{(i)}$. As the algorithm iterates, $p^{(i)}(n,g)$
would converge to a stable distribution satisfying $p^{(i)}(n,g)=0$
or 1.

The EDA algorithm not only preserves the current best individual,
but also introduces diversity by sampling the probability distribution
to generate new individuals, and hence is able to escape from local
optimums with high probability. The EDA algorithm for multi-user pilot
pattern optimization is summarized in Algorithm \ref{alg:AO}.

\begin{algorithm}[t]
{\small{}\caption{\label{alg:AO}The EDA algorithm for multi-user pilot pattern optimization }
}{\small\par}

\textbf{Input:} $Q,T,\mathcal{R}_{s}$, $\beta_{g},$\textcolor{black}{{}
}$\sigma_{g},\rho_{g,1},\rho_{g,2},\forall g$\textcolor{black}{,
maximum iteration number $I$.}

\textbf{Output:} The optimal pilot pattern matrix $\mathbf{W}_{\textrm{best}}^{(I)}$.

\begin{algorithmic}[1]

\STATE Randomly generate $Q$ individuals of the pilot pattern matrix
$\mathbf{W}_{q}^{(0)}$ to initialize the population $\mathcal{W}^{(0)}$.

\FOR{ $i=1,\cdots,I$}

\STATE Evaluate the fitness of individuals $\mathbf{W}_{q}^{(i-1)}$
in $\mathcal{W}^{(i-1)}$ by calculating the objective function of
$\mathcal{P}$.

\STATE Select the best $T$ individuals from $\mathcal{W}^{(i-1)}$
to construct a new population $\mathcal{\widetilde{W}}^{(i)}$. 

\STATE Calculate pilot occurrence probability matrix $\mathbf{P}^{(i)}\in\mathbb{R}^{N\times G}$,
of which the element is calculated by (\ref{eq:Pro}).

\STATE Formulate the conditional Bernoulli probability distribution
$\textrm{Bern}(\mathbf{P}^{(i)})$ with probability $P(\mathbf{W}_{q}^{(i)}(n,g)=1)=p^{(i)}(n,g)$
subject to (\ref{eq:P2_SRL})-(\ref{eq:P2_bandwidth}).

\STATE Sample new individuals $\mathbf{W}_{q}^{(i)}$ from probability
distribution $\textrm{Bern}(\mathbf{P}^{(i)})$, i.e., $\mathbf{W}_{q}^{(i)}(n,g)\sim\textrm{Bern}(p^{(i)}(n,g))$.

\STATE Generate a new population $\mathcal{W}^{(i)}$ with individuals
$\mathbf{W}_{q}^{(i)},q=1,...,Q$.

\STATE Preserve the best individual $\mathbf{W}_{\textrm{best}}^{(i-1)}$
from $\mathcal{W}^{(i-1)}$ to $\mathcal{W}^{(i)}$, i.e., $\mathbf{W}_{1}^{(i)}=\mathbf{W}_{\textrm{best}}^{(i-1)}$.

\ENDFOR

\end{algorithmic}
\end{algorithm}

\section{Extension to Multiband Scenarios\label{sec:Tra}}

Multiband technology is proposed to achieve high-resolution delay
estimation by jointly utilizing measurements from received OFDM signals
across various non-contiguous frequency bands \cite{ESPRIT2}. This
technology aligns with our goal to enhance channel extrapolation performance
and thus motivates the pilot pattern design in multiband scenarios
to further improve delay resolution of the system.

\begin{figure}[t]
\centering{}\includegraphics[width=8cm]{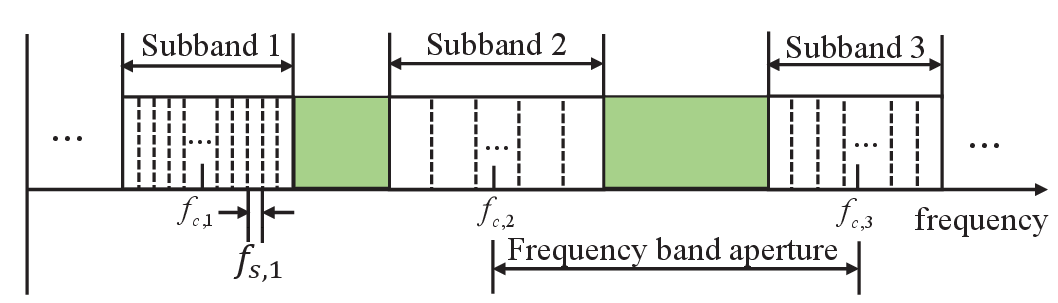}\caption{\label{fig:MBDistri}Frequency distribution of the multiband system
for pilot transmission.}
\end{figure}

As shown in Fig. \ref{fig:MBDistri}, the spectrum resource used for
pilot transmission consists of a number of non-contiguous frequency
subbands in the presence of frequency band apertures, e.g., $f_{c,2}-f_{c,1}$,
with central frequencies $f_{c,m}$ and subcarrier spacing $f_{s,m}$
of the $m$-th subband. The frequency subbands allocated to other
systems or applications are illustrated by the green color and cannot
be utilized for pilot transmission. As discussed in \cite{TSGE_early,ESPRIT2},
the multiband gains originated from the presence of frequency band
apertures in multiband systems can be exploited to improve delay resolution
of the system.

In this section, we extend our proposed multi-user pilot pattern optimization
scheme to the multiband scenarios. We first formulate the multiband
multi-user signal model. Then, we derive the performance metrics ISL
and SRL based on the multiband multi-user signal model. Finally, the
multiband multi-user pilot pattern optimization problem is formulated.

\subsection{Multiband Signal Model}

During the period of a single OFDM symbol, the multiband multi-user
received signal model can be written as \cite{TSGE_early,CS2020}
\begin{equation}
y_{m,n}^{\textrm{MB}}\!=\sum_{g=1}^{G}\sum_{z=1}^{Z}\!\mathbf{h}_{g,z}^{\textrm{MB}}(m,n)\mathbf{w}_{g}^{\textrm{MB}}(m,n)\mathbf{x}_{z}^{\textrm{MB}}(m,n)\!+\!\mathbf{n}_{g,z}^{\textrm{MB}}(m,n),\label{eq:original_signal}
\end{equation}
where
\[
\mathbf{h}_{g,z}^{\textrm{MB}}(m,n)=\sum_{k=1}^{K_{g,z}}\!\alpha_{g,z,k}e^{-j2\pi f_{m,n}\tau_{g,z,k}}e^{-j2\pi nf_{\!s,\!m}\delta_{m}}e^{j\varphi_{m}}
\]
denotes the multiband CFR. The index $(m,n)$ represents the $n$-th
subcarrier of the $m$-th subband, $m=1,...,M$, $n\in\mathcal{N_{\mathit{m}}}\triangleq\left\{ -\frac{N_{m}-1}{2},...,\frac{N_{m}-1}{2}\right\} $,
$f_{m,n}=f_{c,m}+nf_{s,m}$. We assume that the subcarrier number
of the $m$-th subband, $N_{\mathit{m}},\forall m,$ is an odd number
without loss of generality, and denote $N^{\textrm{MB}}=N_{1}+\ldots+N_{M}$
as the number of subcarriers over all subbands. The notations $\mathbf{w}_{g}^{\textrm{MB}}\in\left\{ 0,1\right\} ^{N^{\textrm{MB}}\times1}$
and $\mathbf{x}_{z}^{\textrm{MB}}\in\mathbb{C}^{N^{\textrm{MB}}\text{\texttimes}1}$
denote the multiband pilot pattern vector of the $g$-th group and
the $z$-th multiband SRS sequence, respectively. $\mathbf{n}_{g,z}^{\textrm{MB}}\in\mathbb{C}^{N^{\textrm{MB}}\text{\texttimes}1}$
denotes the AWGN with each element having zero mean and variance $\sigma_{ns,g,z}^{2}$.
Due to the hardware imperfections, multiband systems are affected
by two phase distortion factors, i.e., random phase offset $\varphi_{m}$
and receiver timing offset $\delta_{m}$, which differ from band to
band and need to be calibrated \cite{nonidealfoctor3}.

\subsection{Problem Formulation}

\subsubsection{Derivation of ISL in multiband multi-user scenarios}

In multiband scenarios, the AF of the $(g,z)$-th user can be derived
as
\begin{eqnarray}
 &  & \chi_{g,z}^{\textrm{MB}}(\Delta\tau)\nonumber \\
 &  & =\left(\mathbf{w}_{g}^{\textrm{MB}}\ast\mathbf{x}_{z}^{\textrm{MB}}\ast\boldsymbol{a}^{\textrm{MB}}(\tau_{1})\right)^{H}\left(\mathbf{w}_{g}^{\textrm{MB}}\ast\mathbf{x}_{z}^{\textrm{MB}}\ast\boldsymbol{a}^{\textrm{MB}}(\tau_{2})\right)\nonumber \\
 &  & =(\mathbf{w}_{g}^{\textrm{MB}})^{T}\boldsymbol{a}^{\textrm{MB}}(\Delta\tau),\label{eq:AF-MB}
\end{eqnarray}
where $\boldsymbol{a}^{\textrm{MB}}(\Delta\tau)\in\mathbb{C}^{N^{\textrm{MB}}\text{\texttimes}1}$
denotes the multiband frequency-domain steering vector consisting
of the elements $e^{-j2\pi f_{m,n}\Delta\tau-j2\pi nf_{\!s,\!m}\delta_{m}+\varphi_{m}},\forall n,m.$
Then, substituting (\ref{eq:AF-MB}) into (\ref{eq:ISL}), the multiband
ISL can be written as
\begin{equation}
\textrm{ISL}_{g}^{\textrm{MB}}=\frac{(\mathbf{w}_{g}^{\textrm{MB}})^{T}\mathbf{G}^{\textrm{MB}}\mathbf{w}_{g}^{\textrm{MB}}}{\left|\mathcal{R}_{s}\right|\left((\mathbf{w}_{g}^{\textrm{MB}})^{T}\boldsymbol{1}\right)^{2}},\label{eq:ISL_2-1}
\end{equation}
where
\[
\mathbf{G}^{\textrm{MB}}=\left[\begin{array}{ccc}
\mathbf{G}_{1,1} & \ldots & \mathbf{G}_{1,M}\\
\vdots & \ddots & \vdots\\
\mathbf{G}_{M,1} & \cdots & \mathbf{G}_{M,M}
\end{array}\right],
\]
\[
\mathbf{G}_{i,j}(n,l)=\frac{\sin(2\pi(f_{i,n}-f_{j,l})b)-\sin(2\pi(f_{i,n}-f_{j,l})a)}{\pi(f_{i,n}-f_{j,l})}.
\]
Note that in multiband scenarios, the matrix $\mathbf{G}^{\textrm{MB}}$
no longer preserves the Toeplitz structure in most cases.

\subsubsection{Derivation of SRL in multiband multi-user scenarios }

As compared to the SRL in single-band scenarios, the calculation of
SRL in multi-user multiband scenarios requires the derivation of CRB
based on the multiband signal model (\ref{eq:original_signal}) in
the presence of phase distortion factors. 

To calculate SRL, we firstly need to derive the CRB for the delay
separation $\Delta\tau_{g,z}$, which is given by
\begin{eqnarray}
\textrm{CRB}_{\Delta\tau_{g,z}}^{\textrm{MB}} & = & \frac{\partial\Delta\tau_{g,z}}{\partial\boldsymbol{\eta}_{g,z}}^{T}\mathbf{J}_{\boldsymbol{\eta}_{g,z}}^{-1}\frac{\partial\Delta\tau_{g,z}}{\partial\boldsymbol{\eta}_{g,z}},\label{eq:C_delattau}
\end{eqnarray}
where $\mathbf{J}_{\boldsymbol{\eta}_{g,z}}$ is the FIM associated
with the vector $\boldsymbol{\eta}_{g,z}\triangleq[\boldsymbol{\theta}_{g,z},\boldsymbol{\varphi},\boldsymbol{\delta}]^{T}$,
$\boldsymbol{\varphi}=\left[\varphi_{2},\ldots,\varphi_{M}\right]^{T}$,
$\boldsymbol{\delta}=\left[\delta_{1},\ldots,\delta_{M}\right]^{T}$.
To avoid a singular FIM of $\mathbf{J}_{\boldsymbol{\eta}_{g,z}}$,
we set $f_{c,1}=0,\varphi_{1}=0,$ and assume that $\delta_{m},\forall m$
follows a prior distribution $p\left(\delta_{m}\right)\sim\mathcal{N}\left(0,\sigma_{p}^{2}\right)$
\cite{TSGE_early}. Then, from (\ref{eq:C_delattau}), we have
\[
\textrm{CRB}_{\Delta\tau_{g,z}}^{\textrm{MB}}=\mathbf{J}_{\boldsymbol{\eta}_{g,z}}^{-1}(1,1)+\mathbf{J}_{\boldsymbol{\eta}_{g,z}}^{-1}(2,2)-\mathbf{J}_{\boldsymbol{\eta}_{g,z}}^{-1}(1,2)-\mathbf{J}_{\boldsymbol{\eta}_{g,z}}^{-1}(2,1),
\]
where $\mathbf{J}_{\boldsymbol{\eta}_{g,z}}\triangleq\mathbf{J}_{g,z}^{(w)}+\mathbf{J}_{g,z}^{(p)}$,
$\mathbf{J}_{g,z}^{(w)}$ and $\mathbf{J}_{g,z}^{(p)}$ are the FIMs
from the observations and the a priori knowledge of $\boldsymbol{\delta}$,
respectively, which are defined as \cite{BCRB}
\[
\mathbf{J}_{g,z}^{(w)}\triangleq\mathbb{E}_{\mathbf{y}^{\textrm{MB}},\boldsymbol{\delta}}\left[-\frac{\partial^{2}\ln f(\mathbf{y}^{\textrm{MB}}|\boldsymbol{\eta}_{g,z})}{\partial\boldsymbol{\eta}_{g,z}\partial\boldsymbol{\eta}_{g,z}^{T}}\right],
\]
\[
\mathbf{J}_{g,z}^{(p)}\triangleq\mathbb{E}_{\boldsymbol{\delta}}\left[-\frac{\partial^{2}\ln f(\boldsymbol{\delta})}{\partial\boldsymbol{\eta}_{g,z}\partial\boldsymbol{\eta}_{g,z}^{T}}\right].
\]
The notation $f(\boldsymbol{\delta})\propto\exp\{-\frac{\|\boldsymbol{\delta}\|_{2}^{2}}{2\sigma_{p}^{2}}\}$
denotes the prior distribution of $\boldsymbol{\delta}$. 
\begin{figure*}[tbh]
\begin{equation}
\mathbf{J}_{\boldsymbol{\eta}_{g,z}}=\left[\begin{array}{cccc}
\Psi(\boldsymbol{\tau}_{g,z},\boldsymbol{\tau}_{g,z}) & \Psi(\boldsymbol{\tau}_{g,z},\boldsymbol{\alpha}_{g,z}) & \Psi(\boldsymbol{\tau}_{g,z},\boldsymbol{\varphi}) & \Psi(\boldsymbol{\tau}_{g,z},\boldsymbol{\delta})\\
\Psi(\boldsymbol{\alpha}_{g,z},\boldsymbol{\tau}_{g,z}) & \Psi(\boldsymbol{\alpha}_{g,z},\boldsymbol{\alpha}_{g,z}) & \Psi(\boldsymbol{\alpha}_{g,z},\boldsymbol{\varphi}) & \Psi(\boldsymbol{\alpha}_{g,z},\boldsymbol{\delta})\\
\Psi(\boldsymbol{\varphi},\boldsymbol{\tau}_{g,z}) & \Psi(\boldsymbol{\varphi},\boldsymbol{\alpha}_{g,z}) & \Psi(\boldsymbol{\varphi},\boldsymbol{\varphi}) & \Psi(\boldsymbol{\varphi},\boldsymbol{\delta})\\
\Psi(\boldsymbol{\delta},\boldsymbol{\tau}_{g,z}) & \Psi(\boldsymbol{\delta},\boldsymbol{\alpha}_{g,z}) & \Psi(\boldsymbol{\delta},\boldsymbol{\varphi}) & \Psi(\boldsymbol{\delta},\boldsymbol{\delta})
\end{array}\right].\label{eq:FIM}
\end{equation}
\end{figure*}
Furthermore, $\mathbf{J}_{\boldsymbol{\eta}_{g,z}}$ can be constructed
as in (\ref{eq:FIM}). The entries in $\mathbf{J}_{\boldsymbol{\eta}_{g,z}}$
are derived in Appendix \ref{subsec:AppendixB}.

\subsubsection{Problem formulation in multiband scenarios}

The optimization problem in multiband scenarios can be formulated
as
\begin{eqnarray}
\mathcal{P^{\textrm{MB}}}: & \underset{\mathcal{\varXi}}{\min} & \underset{g}{\max}\left\{ \textrm{ISL}_{g}^{\textrm{MB}}\right\} \nonumber \\
 & \text{s.t. } & \ensuremath{\textrm{SRL}_{g}^{\textrm{MB}}\leq\beta_{g}^{\textrm{MB}},\forall g},\label{eq:P2_SRL-1}\\
 &  & \sum_{g=1}^{G}\mathbf{w}_{g}^{\textrm{MB}}\leq\boldsymbol{1},\label{eq:P2_bandwidth-1}
\end{eqnarray}
where $\varXi\triangleq\left\{ \mathbf{w}_{1}^{\textrm{MB}},...,\mathbf{w}_{G}^{\textrm{MB}}\right\} $
denotes the optimization variables set, $\textrm{SRL}_{g}^{\textrm{MB}}$
denotes the multiband SRL of the $g$-th group, and $\beta_{g}^{\textrm{MB}}$
denotes the upper limit on the multiband SRL of the $g$-th group.
The multi-user multiband pilot optimization problem $\mathcal{P^{\textrm{MB}}}$
can also be optimized in the offline manner using the EDA algorithm,
which is similar to the offline optimization in single-band scenarios.

\section{Simulation Results\label{sec:Simulation-Results}}

In this section, we provide numerical results to validate the effectiveness
of our proposed multi-user pilot pattern optimization scheme. In the
default setup, the OFDM system is equipped with carrier frequency
3.5 GHz, the number of subcarriers $N=256,$ and the subcarrier spacing
$f_{s}=120$ KHz.\textcolor{blue}{{} }The $\textrm{SNR}=15$ dB. We
consider 4 users multiplexing in code-domain and frequency-domain
for $G=2,Z=2$, as depicted in Fig. \ref{fig:Bas_pattern}. Two users
(User 1, User 2) transmit pilot symbols generated from a ZC sequence
in set $\Gamma$ to the BS with the optimized frequency-domain pilot
pattern. The other two users (User 3, User 4) transmit pilot symbols
with the same optimized pilot pattern as User 1 and User 2, but they
employ another ZC sequence from the orthogonality set $\Gamma$. In
EDA algorithm, we set $\rho_{g,1}=\rho_{g,2}=1$, $\sigma_{g}=0.1778,\forall g,$
$Q=400,T=200,$ and $I=60$. The channel samples are generated by
the QuaDRiGa toolbox \cite{Quadriga} according to the 3D-UMa NLOS
model defined by 3GPP R16 specifications \cite{3gpp_Rel16}.

For comparison, we consider the following two benchmark schemes, whose
pilot patterns for 4 users have been shown in Fig. \ref{fig:Bas_pattern}:
\begin{itemize}
\item \textbf{Baseline 1} (Uniform pilot pattern): Each user sequentially
occupies a continuous segment of the frequency band with equivalent
bandwidth. Take 4 users multiplexing for example: One user occupies
the first half of the frequency band with evenly spaced pilot symbols,
while the other user occupies the second half of the frequency band
with evenly spaced pilot symbols, too. The other two users adopt the
same code-domain multiplexing method as the proposed scheme. This
multi-user pilot pattern has been employed in current 5G NR systems.
\item \textbf{Baseline 2} (Random pilot pattern): The users transmit pilot
symbols with a random distribution in subcarriers. The code-domain
multiplexing method is the same as the proposed scheme.
\end{itemize}

\subsection{Single-Band }

\begin{figure}[t]
\centering{}\textcolor{blue}{\includegraphics[width=8cm]{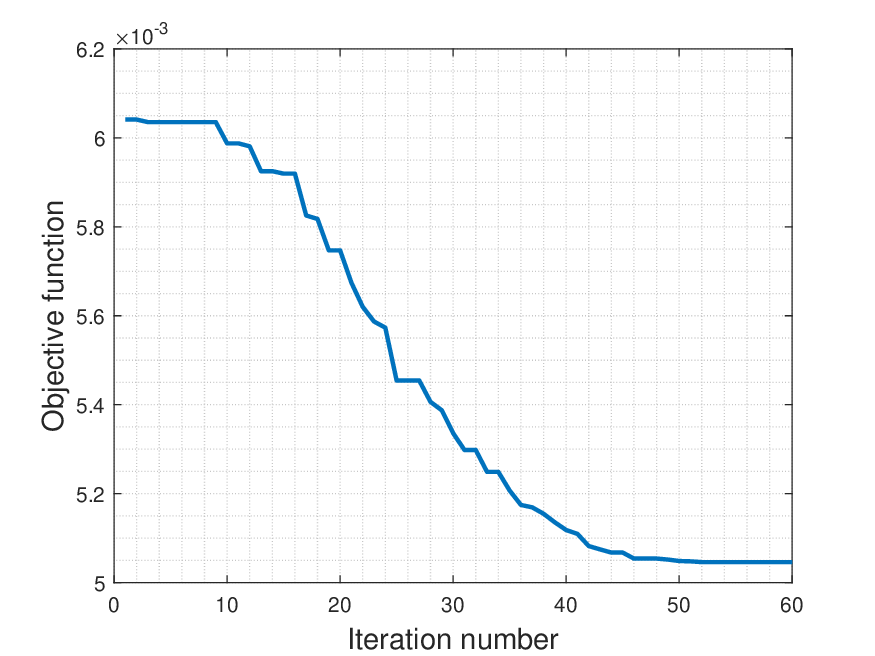}\caption{\label{fig:Convergence-behavior-of}Convergence behavior of the EDA
algorithm in multi-user pilot pattern optimization.}
}
\end{figure}

\begin{figure}[t]
\centering{}\textcolor{blue}{\includegraphics[width=8cm]{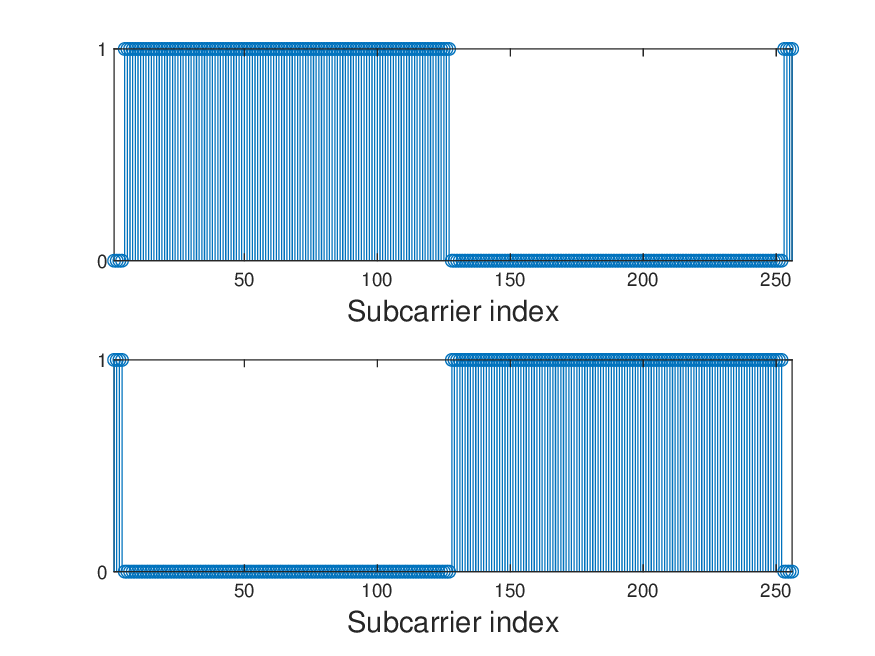}\caption{\label{fig:Pat_C2M1}An illustration of the optimized pilot pattern
of two users for $M=1$.}
}
\end{figure}

We first illustrate the convergence behavior of the adopted EDA algorithm
in multi-user pilot pattern optimization. As illustrated in Fig. \ref{fig:Convergence-behavior-of},
EDA converges within 50 iterations (up to a small convergence error).
The optimized pilot pattern of the two users in frequency domain has
been shown in Fig. \ref{fig:Pat_C2M1}. The pilot pattern for each
user consists of two non-contiguous parts. Each part occupies the
subcarriers with evenly spaced symbols. Besides, one part occupies
the majority of the pilot symbols, while the other part only occupies
the minority of the pilot symbols. This result can be justified that
it takes into account the impact of both the employed metrics, SRL
and ISL, in the optimization process. Specifically, the objective
of minimizing ISL in $\mathcal{P}$ requires the majority of pilot
symbols for each user to be arranged as evenly spaced as possible.
However, due to the SRL constraint in $\mathcal{P}$, the final pilot
pattern cannot have all pilot symbols evenly spaced as Baseline 1
does. It has to separate out a small portion of pilot symbols to span
a larger frequency range to satisfy the SRL requirement.\textcolor{blue}{}
\textcolor{blue}{}

\begin{figure}[t]
\centering{}\textcolor{blue}{\includegraphics[width=8cm]{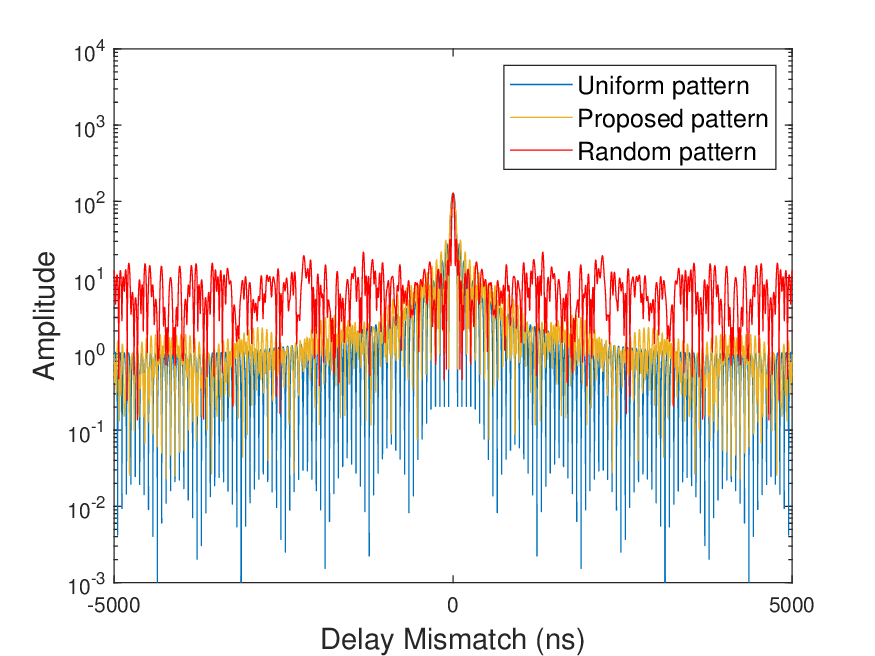}\caption{\label{fig:AF_M1}An illustration of the AFs for $M=1$.}
}
\end{figure}

\begin{figure}[t]
\centering{}\textcolor{blue}{\includegraphics[width=8cm]{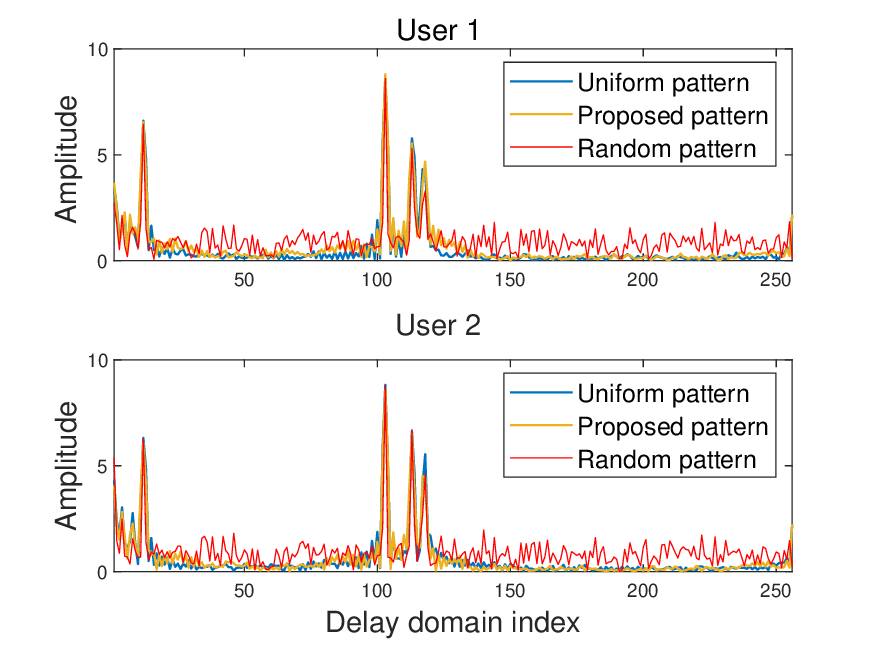}\caption{\label{fig:Delay_Profile}Amplitude delay profiles in delay domain
for $M=1$.}
}
\end{figure}

In Fig. \ref{fig:AF_M1}, we present the AFs with regard to the optimized
pilot pattern and baselines. As can be seen, the side-lobe level based
on our proposed pilot pattern is much lower than that based on the
random pilot pattern and it is nearly close to the side-lobe level
based on the uniform pilot pattern. This phenomenon implies a superior
multi-user interference mitigation capability of our proposed pilot
pattern, as verified in Fig. \ref{fig:Delay_Profile}, where the amplitude
delay profiles are illustrated to show the impact of various pilot
patterns on multi-user interference in code-domain multiplexing. As
can be seen, the level of the multi-user interference of our proposed
scheme is as low as the uniform pilot pattern scheme, and it is much
lower than the random scheme.\textcolor{blue}{}
\textcolor{blue}{}

Then, we compare the delay SRL values of our proposed pilot pattern
scheme with the uniform pilot pattern scheme and the random pilot
pattern scheme, which are 2.8882 ns, 5.7720 ns, and 2.9974 ns, respectively.
It can be seen that our proposed pilot pattern scheme has much better
SRL than the uniform pilot pattern scheme due to the effective delay
SRL constraint (\ref{eq:P2_SRL}) in $\mathcal{P}$.

Finally, we evaluate the channel extrapolation performance of the
proposed pilot pattern and baselines using the particle swarm optimization-least
square (PSO-LS) algorithm \cite{TSGE_early}. PSO-LS has a strong
global search ability to find the optimal solution of a maximum likelihood
(ML) estimation problem. In fact, PSO-LS is proposed for multiband
delay estimation, which needs to solve a multi-dimensional multi-modal
optimization problem that requires a robust global optimization ability.
To have a proper initialization of the PSO-LS algorithm, e.g., determining
the number of MPCs and the search range, the Turbo-CS algorithm has
been employed as the initial phase estimator \cite{TSGE_early}. For
fair comparison, we utilize the PSO-LS algorithm for channel extrapolation
in both single-band and multiband scenarios.

Specifically, we first get the decoupled single-user observations
$\hat{\mathbf{y}}_{g,z}$ based on the multi-user interference cancellation
procedure as depicted in Fig. \ref{fig:FlowChart}. Then, we construct
the ML estimation problem based on $\hat{\mathbf{y}}_{g,z}$ as $\underset{\alpha_{g,z,k},\tau_{g,z,k}}{\min}\left\Vert \hat{\mathbf{y}}_{g,z}-\textrm{diag}\left(\mathbf{w}_{g}^{\textrm{opt}}\right)\mathbf{h}_{g,z}\right\Vert _{2}^{2}$,
where $\mathbf{w}_{g}^{\textrm{opt}}$ denotes the optimized pilot
pattern of the $g$-th group from the output of Algorithm \ref{alg:AO}.
We solve it using PSO-LS to estimate the channel parameters $\alpha_{g,z,k},\tau_{g,z,k}$.
Finally, the full-band channel can be recovered based on the estimated
channel parameters.

\begin{figure}[t]
\centering{}\textcolor{blue}{\includegraphics[width=8cm]{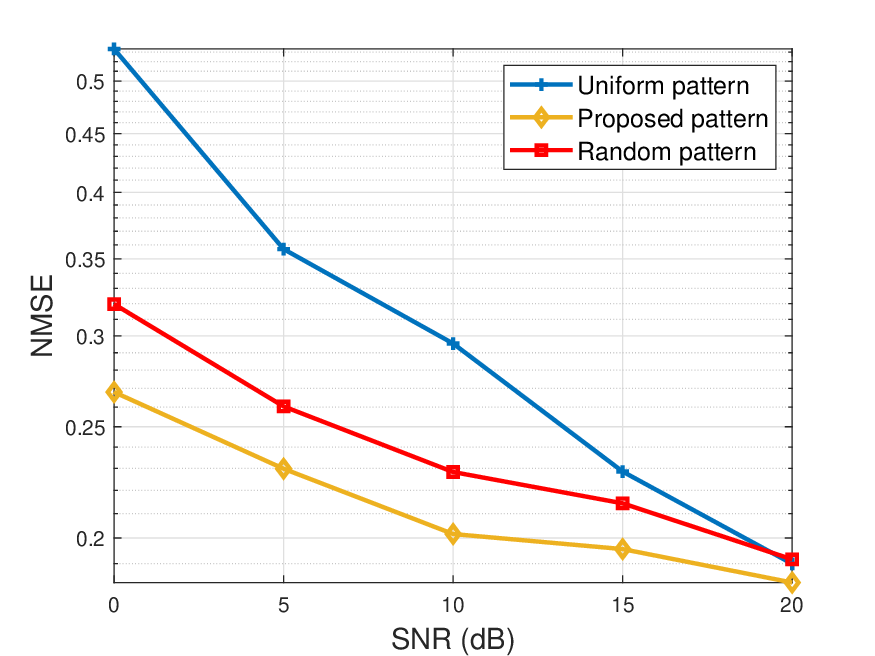}\caption{\label{fig:NMSE_C2M1}NMSE of the full-band channel versus SNR for
$M=1$.}
}
\end{figure}

We choose NMSE as the performance metric to evaluate the channel extrapolation
performance of various schemes, which is defined as
\begin{equation}
\textrm{NMSE}\triangleq\mathbb{E}\left\{ \frac{1}{GZ}\sum_{g=1}^{G}\sum_{z=1}^{Z}\frac{\left\Vert \hat{\mathbf{h}}_{g,z}-\mathbf{h}_{g,z}\right\Vert _{2}^{2}}{\left\Vert \mathbf{h}_{g,z}\right\Vert _{2}^{2}}\right\} ,\label{eq:NMSE}
\end{equation}
where $\hat{\mathbf{h}}_{g,z}$ is the estimate of $\mathbf{h}_{g,z}$,
the expectations in (\ref{eq:NMSE}) are computed by averaging over
500 Monte Carlo trials. As shown in Fig. \ref{fig:NMSE_C2M1}, the
scheme based on the proposed pilot pattern has the lowest NMSE of
channel extrapolation as compared to baselines. Besides, the performance
gap between our proposed scheme and the uniform pilot pattern scheme
is larger in lower SNR region.

In summary, though uniform pilot pattern scheme has strong code-domain
multiplexing capability with effective multi-user interference cancellation,
its drawback lies in its insufficient delay resolution and thus results
in poor channel extrapolation performance. On the other hand, though
random pilot pattern leads to sufficient delay resolution of the system,
but it suffers from severe multi-user interference in code-domain
multiplexing, and due to the high level side-lobe of AF causing significant
delay ambiguity, the accuracy of delay estimation is limited. In contrast
to these two baselines, our proposed pilot pattern scheme has both
the advantages of robust resistance to multi-user interference and
the high resolution delay estimation. Consequently, the overall channel
extrapolation performance of the proposed scheme is better than that
of the baseline schemes. 

\textcolor{blue}{}

\subsection{Multiband}

In multiband scenarios of the default setup, the OFDM system is equipped
with carrier frequency $f_{c,1}=3.5$ GHz, $f_{c,2}=3.9$ GHz, the
number of subcarriers $N_{1}=N_{2}=128$, and the subcarrier spacing
$f_{s,1}=f_{s,2}=120$ KHz at $M=2$ subbands. We consider 2 users
($G=2,Z=2$) and 3 users ($G=3,Z=2$) multiplexing in frequency-domain,
respectively, and for the case of $G=3$, we set $N_{1}=N_{2}=192$.

\begin{figure}[t]
\begin{centering}
\begin{minipage}[t]{0.45\textwidth}%
\begin{center}
\subfloat[\label{fig:Pat_M2C2}]{\begin{centering}
\includegraphics[width=80mm]{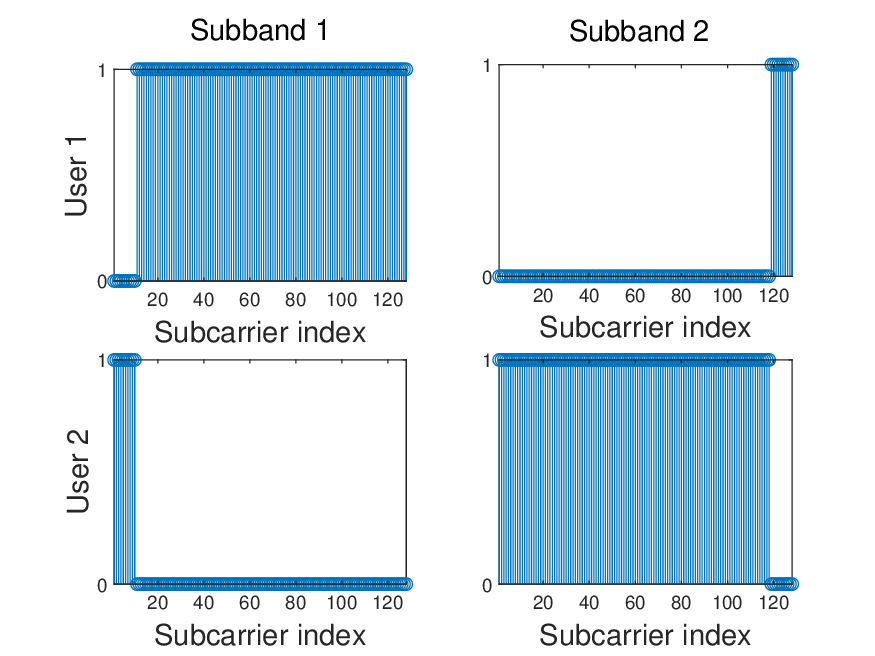}
\par\end{centering}
}
\par\end{center}%
\end{minipage}\hfill{}%
\begin{minipage}[t]{0.45\textwidth}%
\begin{center}
\subfloat[\label{fig:Pat_M2C3}]{\begin{centering}
\includegraphics[width=8cm]{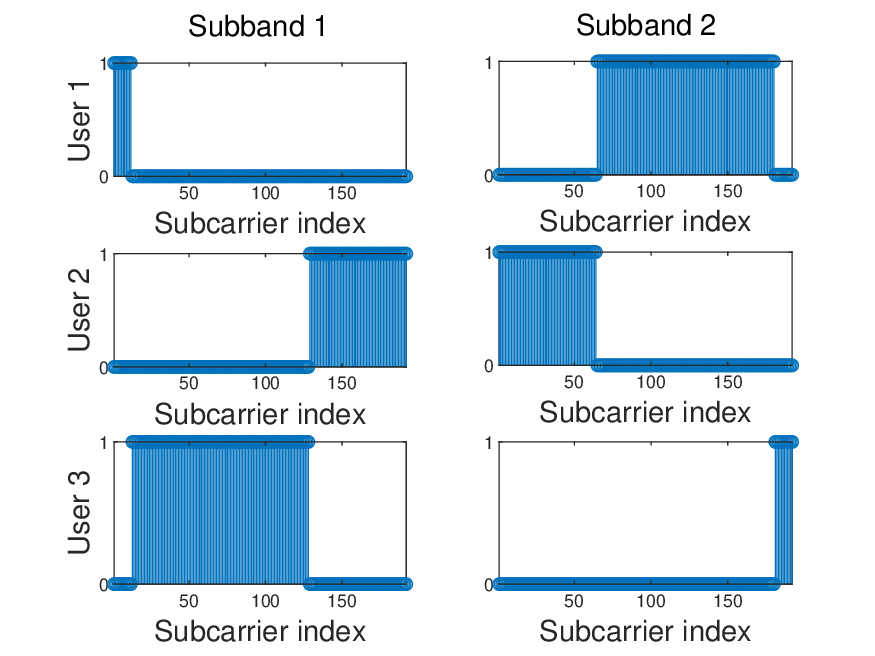}
\par\end{centering}
}
\par\end{center}%
\end{minipage}
\par\end{centering}
\medskip{}

\centering{}\caption{\label{fig:Pat_M2}An illustration of the optimized pilot pattern
for $M=2$: (a) $G=2$; (b) $G=3$.}
\end{figure}

The optimized pilot pattern has been shown in Fig. \ref{fig:Pat_M2C2},
where the pilot symbols of each user are arranged in both subbands.
Moreover, in each subband, the pilot symbols are evenly spaced. This
pilot pattern is rational since it ensures the majority of pilot symbols
evenly spaced to reduce the multi-user interference and simultaneously
spans two subbands to reserve the frequency band aperture structure
(which is the source of the multiband gains) to improve delay resolution
of the system. In Fig. \ref{fig:Pat_M2C3}, we can see that all three
users have arranged the pilot symbols into two subbands, and in each
subband, the pilot symbols are evenly spaced as the case $G=2$ does. 

\begin{figure}[t]
\centering{}\textcolor{blue}{\includegraphics[width=8cm]{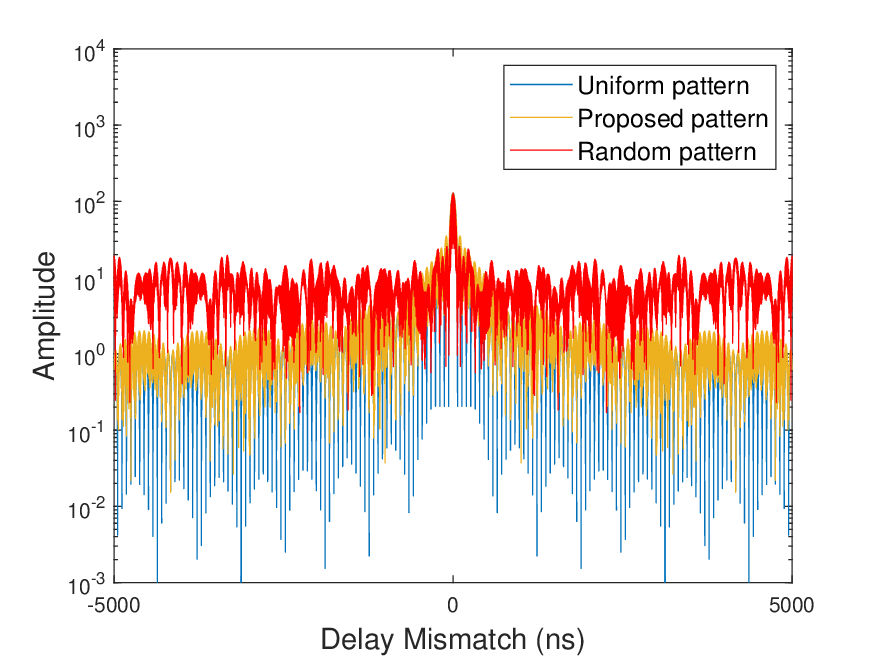}\caption{\label{fig:AF_M2}An illustration of the AFs for $M=2$.}
}
\end{figure}

We plot the curve of AF in Fig. \ref{fig:AF_M2}. It can be seen that
the the proposed pilot pattern still achieves a lower side-lobe level
than the random pilot pattern as that in single-band scenarios. Moreover,
in multiband scenarios, the side-lobe fluctuates more frequently compared
with that in single-band scenarios. It can be justified that in the
multiband AF expression (\ref{eq:AF-MB}), the delay mismatch $\Delta\tau$
is multiplied by a high carrier frequency $f_{c,m}$. Consequently,
a slight variation of $\Delta\tau$ may lead to a significant phase
shift of $\boldsymbol{a}^{\textrm{MB}}(\Delta\tau)$ and finally it
is reflected in the violent oscillation of the AF. \textcolor{blue}{}
\begin{figure}[t]
\begin{centering}
\begin{minipage}[t]{0.45\textwidth}%
\begin{center}
\subfloat[\label{fig:NMSE_MB_C4}]{\begin{centering}
\includegraphics[width=80mm]{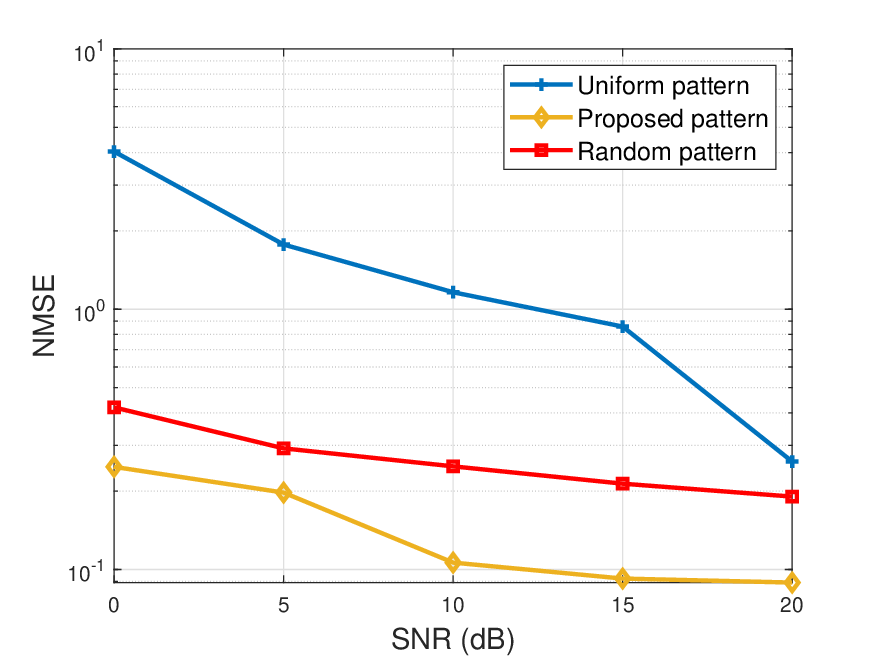}
\par\end{centering}
}
\par\end{center}%
\end{minipage}\hfill{}%
\begin{minipage}[t]{0.45\textwidth}%
\begin{center}
\subfloat[\label{fig:NMSE_MB_C6}]{\begin{raggedright}
\includegraphics[width=8cm]{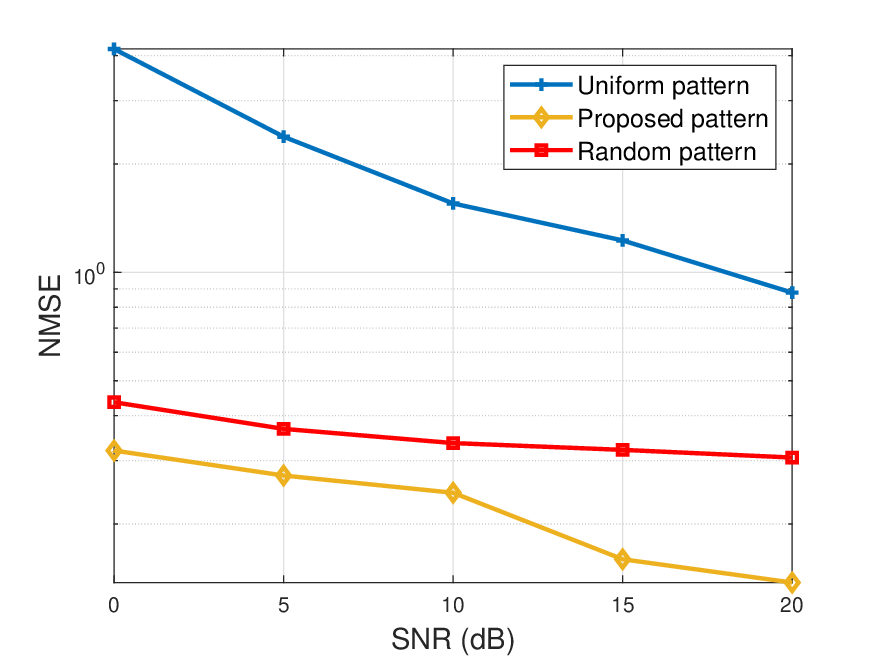}
\par\end{raggedright}
}
\par\end{center}%
\end{minipage}
\par\end{centering}
\medskip{}

\centering{}\caption{\label{fig:NMSE_MB}NMSE of the full-band channel versus SNR for $M=2$:
(a) $G=2$; (b) $G=3$.}
\end{figure}

Then, we calculate the delay SRL values of our proposed pilot pattern
scheme, the uniform pilot pattern scheme, and the random pilot pattern
scheme for the case of $G=2$, which are 0.5707 ns, 5.7977 ns, and
0.5844 ns, respectively. It can be seen that the SRL of our proposed
pilot pattern scheme and the random pilot pattern scheme in multiband
scenarios is much better than that in single-band scenarios due to
the multiband gains brought by the inherent frequency band apertures
in multiband systems. The uniform pilot pattern scheme, however, does
not reserve the frequency band aperture structure since the pilots
of each user only occupy a single contiguous subband. Consequently,
the SRL of the uniform pilot pattern scheme has not been significantly
improved. In Fig. \ref{fig:NMSE_MB}, we investigate the extrapolation
performance of the proposed pilot pattern scheme. As can be seen,
in multiband scenarios, both the schemes based on the random pilot
pattern and the proposed pilot pattern have a lower NMSE than the
uniform pilot pattern scheme. As stated of the delay SRL comparison
earlier, the system based on these two pilot patterns has sufficient
delay resolution owing to the exploitation of multiband gains. The
scheme based on the uniform pilot pattern, however, does not reserve
the frequency band aperture structure for any user and thus cannot
exploit the multiband gains to improve delay resolution of the system.

Furthermore, compare the NMSE performance of our proposed scheme in
Fig. \ref{fig:NMSE_MB_C4} to that in Fig. \ref{fig:NMSE_C2M1}, it
can be seen that the channel extrapolation performance of the proposed
pilot pattern scheme is better in multiband scenarios than in single-band
scenarios, which validates the effective leverage of the multiband
gains to improve the channel extrapolation performance by the proposed
pilot pattern scheme.

\section{Conclusion\label{sec:Conclusion}}

In this paper, we have designed a multi-user pilot pattern for channel
extrapolation in OFDM systems. We have formulated an offline multi-user
pilot pattern optimization problem with the objective of minimizing
the maximum ISL among users. By doing this, the multi-user interference
can be well mitigated in code-domain and the side-lobe of the AF is
significantly suppressed. Besides, we have considered a constraint
of the SRL of the system to ensure that it has sufficient delay resolution
for channel extrapolation. Furthermore, we have extended our pilot
pattern optimization scheme to a multiband scenario for the sake of
exploiting the multiband gain to improve delay resolution of the system.
Finally, simulation results show that our proposed pilot pattern outperforms
the baselines for channel extrapolation performance in both single-band
and multiband scenarios. The optimized pilot patterns provide useful
guidance for the design of multi-user pilot pattern in 5G NR systems.\vspace{-0.1in}

\appendix

\subsection{\label{subsec:AppexA}Derivation of $\mathbf{J}_{\boldsymbol{\theta}_{g,z}}$}

The FIM $\mathbf{J}_{\boldsymbol{\theta}_{g,z}}$ can be structured
as
\begin{equation}
\mathbf{J}_{\boldsymbol{\theta}_{g,z}}=\left[\begin{array}{cc}
\Phi(\boldsymbol{\tau}_{g,z},\boldsymbol{\tau}_{g,z}) & \Phi(\boldsymbol{\tau}_{g,z},\boldsymbol{\alpha}_{g,z})\\
\Phi(\boldsymbol{\alpha}_{g,z},\boldsymbol{\tau}_{g,z}) & \Phi(\boldsymbol{\alpha}_{g,z},\boldsymbol{\alpha}_{g,z})
\end{array}\right].\label{eq:FIM-1}
\end{equation}

For easy of illustration, we omit the subscript $(g,z)$ in the expression
of the entries in $\mathbf{J}_{\boldsymbol{\theta}_{g,z}}$. The expression
is given by
\[
\begin{aligned} & \Phi\left(\tau_{r},\tau_{s}\right)\\
 & \,\,=\frac{8\pi^{2}}{\sigma_{e}^{2}}\sum_{n}(nf_{s})^{2}\textrm{Re}\left\{ (\alpha_{r}^{\prime})^{*}\alpha_{s}^{\prime}e^{j2\pi nf_{s}\left(\tau_{r}-\tau_{s}\right)}\mathbf{w}(n)\right\} ,\\
 & \Phi\left(\tau_{r},\alpha_{s}^{R}\right)=\Phi\left(\alpha_{s}^{R},\tau_{r}\right)\\
 & \,\,=\frac{4\pi}{\sigma_{e}^{2}}\sum_{n}\textrm{Re}\left\{ jnf_{s}(\alpha_{r}^{\prime})^{*}e^{j2\pi nf_{s}\left(\tau_{r}-\tau_{s}\right)}\mathbf{w}(n)\right\} ,\\
 & \Phi\left(\tau_{r},\alpha_{s}^{I}\right)=\Phi\left(\alpha_{s}^{I},\tau_{r}\right),\\
 & \,\,=-\frac{4\pi}{\sigma_{e}^{2}}\sum_{n}\textrm{Re}\left\{ nf_{s}(\alpha_{r}^{\prime})^{*}e^{j2\pi nf_{s}\left(\tau_{r}-\tau_{s}\right)}\mathbf{w}(n)\right\} \\
 & \Phi\!\left(\alpha_{r}^{R},\!\alpha_{s}^{R}\right)\!=\!\Phi\!\left(\!\alpha_{r}^{I},\!\alpha_{s}^{I}\!\right)\!=\!\frac{2}{\sigma_{e}^{2}}\!\sum_{n}\!\cos(2\pi nf_{s}(\tau_{r}\!-\!\tau_{s}))\mathbf{w}(n),\\
 & \Phi\!\left(\alpha_{r}^{R},\!\alpha_{s}^{I}\right)\!=\!-\!\Phi\!\left(\!\alpha_{r}^{I},\!\alpha_{s}^{R}\!\right)\!=\!\frac{2}{\sigma_{e}^{2}}\!\sum_{n}\!\sin(2\pi nf_{s}(\tau_{s}\!-\!\tau_{r}))\mathbf{w}(n),
\end{aligned}
\]
where $r=1,2,s=1,2.$

\subsection{Entries in $\mathbf{J}_{\boldsymbol{\eta}_{g,z}}$\label{subsec:AppendixB}}

For easy of illustration, we also omit the subscript $(g,z)$ in the
expression of the entries, which are written as
\[
\begin{aligned} & \Psi\left(\tau_{r},\tau_{s}\right)\\
 & \,\,=\frac{8\pi^{2}}{\sigma_{ns}^{2}}\sum_{m,n}f_{m,n}^{2}\textrm{Re}\left\{ (\alpha_{r}^{\prime})^{*}\alpha_{s}^{\prime}e^{j2\pi f_{m,n}\left(\tau_{r}-\tau_{s}\right)}\mathbf{w}^{\textrm{MB}}(m,n)\right\} ,\\
 & \Psi\left(\tau_{r},\alpha_{s}^{R}\right)\\
 & \,\,=\frac{4\pi}{\sigma_{ns}^{2}}\sum_{n,m}\textrm{Re}\left\{ jf_{m,n}(\alpha_{r}^{\prime})^{*}e^{j2\pi f_{m,n}\left(\tau_{r}-\tau_{s}\right)}\mathbf{w}^{\textrm{MB}}(m,n)\right\} ,\\
 & \Psi\left(\tau_{r},\alpha_{s}^{I}\right)\\
 & =-\frac{4\pi}{\sigma_{ns}^{2}}\sum_{n,m}\textrm{Re}\left\{ f_{m,n}(\alpha_{r}^{\prime})^{*}e^{j2\pi f_{m,n}\left(\tau_{r}-\tau_{s}\right)}\mathbf{w}^{\textrm{MB}}(m,n)\right\} ,\\
 & \Psi\left(\tau_{r},\varphi_{i}^{\prime}\right)\\
 & =\!-\frac{4\pi}{\sigma_{ns}^{2}}\!\sum_{n,m=i}\!\!\!\textrm{Re}\!\left\{ \!f_{i,n}(\alpha_{r}^{\prime})^{*}\!\!\sum_{k=1}^{K}\alpha_{k}e^{j2\pi f_{i,n}\left(\tau_{r}-\tau_{k}\right)}\mathbf{w}^{\textrm{MB}}(m,n)\!\right\} \!,
\end{aligned}
\]
\[
\begin{aligned} & \Psi\left(\tau_{r},\delta_{i}\right)\\
 & \!=\!\frac{8\pi^{2}}{\sigma_{ns}^{2}}\!\!\sum_{n,m=i}\!\!\!\!nf_{i,n}f_{s,i}\textrm{Re}\!\left\{ \!\!(\alpha_{r}^{\prime})^{*}\!\!\sum_{k=1}^{K}\!\alpha_{k}^{\prime}e^{j2\pi f_{i,n}\left(\!\tau_{r}-\tau_{k}\!\right)}\mathbf{w}^{\textrm{MB}}(m,n)\!\!\right\} \!,\\
 & \Psi\left(\alpha_{r}^{R},\alpha_{s}^{R}\right)=\Psi\left(\alpha_{r}^{I},\alpha_{s}^{I}\right)\\
 & \,\,=\!\frac{2}{\sigma_{ns}^{2}}\!\sum_{m,n}\!\cos(2\pi f_{m,n}(\tau_{r}\!-\!\tau_{s}))\mathbf{w}^{\textrm{MB}}(m,n),\\
 & \Psi\left(\alpha_{r}^{R},\alpha_{s}^{I}\right)=-\Psi\left(\alpha_{r}^{I},\alpha_{s}^{R}\right)\\
 & \,\,=\frac{2}{\sigma_{ns}^{2}}\!\sum_{m,n}\!\sin(2\pi f_{m,n}(\tau_{s}\!-\!\tau_{r}))\mathbf{w}^{\textrm{MB}}(m,n),
\end{aligned}
\]
\[
\begin{aligned} & \Psi\left(\alpha_{r}^{R},\varphi_{i}^{\prime}\right)\\
 & \,\,=\frac{2}{\sigma_{ns}^{2}}\!\sum_{n,m=i}\!\textrm{Re}\left\{ j\sum_{k=1}^{K}\alpha_{k}^{\prime}e^{j2\pi f_{i,n}\left(\tau_{r}-\tau_{k}\right)}\mathbf{w}^{\textrm{MB}}(m,n)\right\} ,\\
 & \Psi\left(\alpha_{r}^{R},\delta_{i}\right)\\
 & =\!-\frac{4\pi}{\sigma_{ns}^{2}}\!\sum_{n,m=i}\!\textrm{Re}\!\left\{ \!jnf_{s,i}\sum_{k=1}^{K}\alpha_{k}^{\prime}e^{j2\pi f_{n,i}\left(\tau_{r}-\tau_{k}\right)}\!\mathbf{w}^{\textrm{MB}}(m,n)\right\} \!,
\end{aligned}
\]
\[
\begin{aligned} & \Psi\left(\alpha_{r}^{I},\varphi_{i}\right)\\
 & \,\,=\frac{2}{\sigma_{ns}^{2}}\sum_{n,m=i}\textrm{Re}\left\{ \sum_{k=1}^{K}\alpha_{k}^{\prime}e^{j2\pi f_{i,n}\left(\tau_{r}-\tau_{k}\right)}\mathbf{w}^{\textrm{MB}}(m,n)\right\} ,\\
 & \Psi\left(\alpha_{r}^{I},\delta_{i}\right)\\
 & =\!-\frac{4\pi}{\sigma_{ns}^{2}}\!\sum_{n,m=i}\!\textrm{Re}\left\{ \!nf_{s,i}\sum_{k=1}^{K}\alpha_{k}^{\prime}e^{j2\pi f_{i,n}\left(\tau_{r}-\tau_{k}\right)}\mathbf{w}^{\textrm{MB}}(m,n)\right\} \!,\\
 & \Psi\left(\varphi_{r}^{\prime},\varphi_{s}^{\prime}\right)=\left\{ \!\begin{array}{c}
\frac{2}{\sigma_{ns}^{2}}\!\underset{n,m=r}{\sum}\left|\sum_{k=1}^{K}\alpha_{k}^{\prime}e^{-j2\pi f_{r,n}\tau_{k}}\right|^{2}\\
\times\mathbf{w}^{\textrm{MB}}(m,n),r=s,\\
0,\text{otherwise},
\end{array}\right.\\
 & \Psi\left(\varphi_{r}^{\prime},\delta_{s}\right)=\left\{ \!\!\!\begin{array}{c}
-\frac{4\pi}{\sigma_{ns}^{2}}\!\underset{n,m=r}{\sum}\!\!nf_{s,r}\!\left|\sum_{k=1}^{K}\alpha_{k}^{\prime}e^{-j2\pi f_{r,n}\tau_{k}}\right|^{2}\\
\times\mathbf{w}^{\textrm{MB}}(m,n),r=s,\\
0,\text{otherwise},
\end{array}\right.
\end{aligned}
\]
\begin{equation}
\begin{aligned} & \Psi\left(\delta_{r},\delta_{s}\right)=\left\{ \!\!\!\begin{array}{c}
\frac{8\pi^{2}}{\sigma_{ns}^{2}}\!\underset{n,m=r}{\sum}\!\!\!\!n^{2}f_{s,r}^{2}\!\left|\sum_{k=1}^{K}\!\alpha_{k}^{\prime}e^{-j2\pi f_{r,n}\tau_{k}}\right|^{2}\\
\times\mathbf{w}^{\textrm{MB}}(m,n)\!\!\!+\!\frac{1}{\sigma_{p}^{2}}\!,r\!=\!s,\\
0,\text{otherwise}.
\end{array}\right.\end{aligned}
\label{eq:FIM4}
\end{equation}


\end{document}